\documentclass[12pt,a4paper,english]{article}

\usepackage[latin1]{inputenc}
\usepackage[centertags]{amsmath}
\usepackage{color}
\usepackage{newlfont}
\usepackage{graphicx}
\usepackage{babel,amsmath,amssymb,amsbsy,amsfonts,latexsym,amsthm,mathrsfs,color}

\definecolor{mycolorred}{rgb}{1, 0, 0}

\newtheorem{thm}{Theorem}[section]

\newtheorem{lemma}[thm]{Lemma}

\newtheorem{prop}[thm]{Proposition}

\newtheorem{remark}[thm]{Remark}

\newtheorem{i-thm}{Theorem}

\newtheorem{i-prop}[i-thm]{Proposition}

\def\acc{\`}
\def \virg{`}

\newcommand{\beq}{\begin{equation}}
\newcommand{\eeq}{\end{equation}}
\def\<{\langle}
\def\>{\rangle}

\def\N{\mathbb{N}}      
\def\R{{\mathbb R}}
\def\E{{\mathbb E}}
\def\P{{\mathbb P}}
\def\D{{\mathbb D}}
\def\L{{\mathbb L}}

\def\H{{\cal H}}
\def\I{\textbf 1}

\def\={{\,=\,}}
\def\+{{\,+\,}}
\def\^gamma{{\tilde{\gamma}}}

\def\cl#1{{\mathscr #1}}

\newcommand{\cvd}{$\quad\Box $
                  \medskip

                 }
\begin{document}

\title{\textbf{Power series representations\\
for European option prices\\
under stochastic volatility models}}

\author{Lucia Caramellino\thanks{Corresponding author. Dipartimento di Matematica, Universit\`a di Roma \emph{Tor Vergata}, Via della Ricerca Scientifica 1, I-00133 Roma, Italy; \texttt{caramell@mat.uniroma2.it}}
\and Giorgio Ferrari\thanks{Dipartimento di Metodi e Modelli per l'Economia, il Territorio e la Finanza, Università di Roma \emph{La sapienza}, Via del Castro Laurenziano 9, 00161 Roma, Italy; \texttt{giorgio.ferrari@uniroma1.it}}
 \and  Roberta Piersimoni\thanks{Dipartimento di Matematica, Universit\`a di Roma \emph{Tor Vergata}, Via della Ricerca Scientifica 1, I-00133 Roma, Italy; \texttt{epiersi@inwind.it}}}
\date{}
\maketitle

{\textbf{Abstract.}} In the context of stochastic volatility models, we study representation formulas in terms of expectations for the power series' coefficients associated to the call price-function. As in Antonelli and Scarlatti \cite{ScarlattiAntonelli} the expansion is done w.r.t. the correlation between the noises driving the underlying asset price process and the volatility process.
We first obtain expressions for the power series' coefficients from the generalized Hull and White formula obtained in Al\`os \cite{Alos1}. Afterwards, we provide representations turning out from the approach for the sensitivity problem tackled by Malliavin calculus techniques, and these allow to handle not only vanilla options. Finally, we show the numerical performance of the associated Monte Carlo estimators for several stochastic volatility models.

\smallskip


\smallskip


\smallskip

{\textbf{JEL} classification}: C02
, G13
, C63
.

\smallskip


\section{Introduction}
In the last three decades a lot of works have been produced in order to generalize the seminal model by Black and Scholes \cite{BS}.
The purpose of many researchers was to formalize models with a more general stock price dynamics able to explain the features observed in the financial market.
Indeed, with the assumption of constant volatility (as in the Black-Scholes model) one cannot explain well known stylized facts like the so called smile effect, volatility clustering, fat tails or excess kurtosis in the log-returns' distribution. It follows that the empirical observation of such phenomena leads to model the volatility itself as a stochastic process with a dynamics correlated with that of the risky asset. The most popular stochastic volatility models are, among others, the models by Hull and White \cite{HullandWhite}, Stein and Stein \cite{bib:stein} or Heston \cite{heston}. We also refer e.g. to Frey \cite{Frey}, Ghysels, Harvey and Renault \cite{GHR} or
Fouque, Papanicolaou and Sircar \cite{bib:fps} for general surveys.

In the context of stochastic volatility models, several recent papers deal with the behavior of the price of derivatives w.r.t. a small parameter. For example, Benhamou, Gobet and Miri \cite{bib:bgm} derive analytical formulas for pricing vanilla options for a time dependent Heston model by using a small volatility of volatility expansion. Another parameter of interest is the correlation between the noises driving the underlying asset price process and the volatility process. From the analysis of financial data it is in fact evident that there exists a negative instantaneous correlation between the volatility and the stock price process. This is exactly what we are interested in: we study here power series representations for European option prices.

We assume (see Section \ref{sect-start} for mathematical details) that the risky asset $S$ and the volatility process $v$ fulfill the dynamics
\begin{align*}
v_s&=v+\int_t^s\mu(v_\theta)\,d\theta+\int_t^s\eta(v_\theta)\, dB_\theta^1\\
S_s&=e^x+\int_t^s rS_\theta\, d\theta+\int_t^sf(v_\theta)S_\theta\big(\rho\,dB_\theta^1+\sqrt{1-\rho^2}\, dB_\theta^2\big) \end{align*}
where $B$ is a $2$-dimensional Brownian motion and $\rho \in (-1,1)$ denotes the correlation between the Brownian motion $B^1$ and the Brownian motion $Z= \rho B^1 + \sqrt{1-\rho^2}B^2$.
In this setting, the well known formula for the call price function is given by
$$
u(t,x,v;\rho)=\mathbb{E}\left(e^{-r(T-t)}\left(S_T^{t,x,v}-K\right)_+\right).
$$
Antonelli and Scarlatti proved in \cite{ScarlattiAntonelli} that under suitable regularity assumptions on $f$, $\eta$ and $\mu$ (that we will recall in Section \ref{sect-as}), $u$
is $C^{\infty}$ as a function of $\rho$ in a neighborhood of $\rho=0$, where it can be developed in power series: there exists $R\in(0,1)$ such that
$$
u(t,x,v;\rho)=\sum\limits_{k\ge 0}g_k(t,x,v)\, \rho^k,\quad |\rho|<R.
$$
The power series coefficients
$$
g_k(t,x,v)=\frac 1 {k!}\partial^k_\rho u(t,x,v;\rho)_{|\rho=0}
$$
solve suitable parabolic partial differential equations and thanks to Feynman-Kac type formulas, they can be represented as expectations of functionals of the diffusion pair $(S,v)$ evaluated at $\rho=0$.

\smallskip

Starting from this fact, in this paper we study alternative probabilistic representations, as an expectation, for the Taylor coefficients $g_k$'s.

\smallskip

We first deal with the results in Al\`os \cite{Alos1}, that concern a generalization of the classical Hull and White formula \cite{HullandWhite} for European option prices.
Starting from the Al\`os formula, we first give an alternative representation for call price function $u$ and then we perform derivatives, getting an expression for the Taylor coefficients with a direct approach which leads to simple formulas.

But from the financial point of view, it is also clear that the Taylor coefficients $g_k$'s can be thought as sensitivities of the price-function $u(t,x,v;\rho)$ with respect to the correlation coefficient $\rho$.
Thus, we obtain probabilistic representations for the $g_k$'s in terms of expectations by means of Malliavin calculus techniques. And this brings to further alternative representation formulas. Here, we get the results by using the Malliavin calculus in the direction of the noise $B^2$. This allows one to find weights which are easy to handle and  worth to set up plain Monte Carlo estimators for the power series coefficients. It is worth to be said that one could actually find results from the use of both the noises $B^1$ and $B^2$ but this procedure brings to difficult formulas, unfeasible to be used in practice.
Moreover, the Malliavin approach does not depend on the particular form of the payoff function, although it must depend on the asset price at maturity, and this is a basic difference with the result in Antonelli and Scarlatti \cite{ScarlattiAntonelli} (even if their result can be generalized to other situations, see Antonelli, Ramponi and Scarlatti \cite{bib:ars}).
So, we numerically study the behavior of the Monte Carlo estimators to price call options by using the Taylor expansion up to the first and the second order, by means of the  representation formulas for the coefficients we provide. It is worth to be said that all representations work efficiently, the Malliavin weights being the ones giving results a little bit less accurate but, at the same time, providing the most flexible approach.

\smallskip

Plan of the paper. Section \ref{sect-start} is devoted to set up the model, the notations and the main requirements we need and Section \ref{sect-as} contains the results obtained by Antonelli and Scarlatti \cite{ScarlattiAntonelli}. In Section \ref{sect-alos} we study the series expansion from the Al\`os \cite{Alos1} generalized Hull and White formula whereas the aim of Section \ref{sect-mall} is to obtain an expression for the coefficient of the series expansion by using the Malliavin calculus approach for the sensitivity problem. Finally, numerical results for several well known stochastic volatility models are discussed in Section \ref{sect-numerics}.

\section{The framework}
\label{sect-start}

We resume here the notations, the model and the requirements we need throughout this paper.

\smallskip

Let $T>0$ be a fixed time horizon and let $(\Omega, \mathcal{F}, \mathbb{P})$ denote a complete probability space on which a $2$-dimensional Brownian motion $B$ 
is defined over $[0,T]$, and we let $(\mathcal{F}_t)_{t\in[0,T]}$ stand for the natural Brownian filtration augmented with $\mathbb{P}$-null sets.

We consider a market model with one risky asset and one risk-free asset (the money market), the latter being assumed to be modeled by means of the instantaneous interest rate $r$.
We assume that the price $S$ of the risky asset and the stochastic volatility $v$ are modeled as follows: for $t\in[0,T]$,  $x\in\mathbb{R}$ and $v>0$, the pair $(v,S)$ evolves as
\begin{align}
v_s&=v+\int_t^s\mu(v_u)\,du+\int_t^s\eta(v_u)\, dB_u^1\label{din2}\\
S_s&=e^x+\int_t^s rS_u\, du+\int_t^sf(v_u)S_u\big(\rho\,dB_u^1+\sqrt{1-\rho^2}\, dB_u^2\big) \label{din1}
\end{align}
for $s\in[t,T]$, where $\rho \in (-1,1)$ denotes the correlation between the Brownian motion $B^1_t$ and the Brownian motion $Z_t= \rho B^1_t + \sqrt{1-\rho^2}B^2_t$.
Thus, (\ref{din2}) and (\ref{din1}) both say that the underlying probability measure $\P$ is assumed to be a risk-neutral one.

\smallskip

Let us consider the following assumption:
\begin{itemize}
\item[\textbf{(H)}] \emph{$f,\mu$,$\eta:\mathbb{R}\to\mathbb{R}$ are $C^{\infty}$ functions whose derivatives of order $\geq 1$ are bounded.}
\end{itemize}
In particular, \textbf{(H)} ensures the existence and uniqueness of a strong solution of (\ref{din2})-(\ref{din1}) and
the processes $\xi_s=\ln S_s$ solves
\begin{equation}\label{din-xi}
\xi_s= x+\int_t^s\Big(r-\frac12f^2(v_u)\Big)du+\int_t^sf(v_u)\big(\rho\,dB_u^1+\sqrt{1-\rho^2}\, dB_u^2\big).
\end{equation}
In what follows, it will be useful to give a special denomination to the process $\xi$ whenever $\rho=0$. Therefore, we set
\begin{equation}\label{din-xi-0}
\hat \xi_s={\xi_s}{|_{\rho=0}}:\quad
\hat \xi_s= x+\int_t^s\Big(r-\frac12f^2(v_u)\Big)du+\int_t^sf(v_u) dB_u^2.
\end{equation}
We will use the notation $\xi_s^{t,x,v}$, $\hat\xi_s^{t,x,v}$ and $v_s^{t,v}$ to stress the dependence of $\xi_s$, $\hat\xi_s$ and $v_s$ w.r.t. the starting instant and position, and this will be done for all the diffusion processes we are going to define.
For further use,
let us also recall that under \textbf{(H)} the first variation process $Y_s=\partial_v v_s^{t,v}$ and its inverse $Z_s=Y_s^{-1}$, $s\geq t$, are both well defined and solve the equations
\begin{align}
Y_s&=1+\int_t^s\mu'(v^{t,v}_u)Y_udu+\int_t^s\eta'(v^{t,v}_u)Y_udB^1_u\label{dv}\\
Z_s&=1+\int_t^s[-\mu'+(\eta')^2](v^{t,v}_u)
Z_udu-\int_t^s\eta'(v^{t,v}_u)Z_udB^1_u\nonumber
\end{align}
respectively.
Equivalently,
$$
Y_s=\exp\Big(\int_t^s\Big[\mu'-\frac 12 (\eta')^2\Big]
(v^{t,v}_u)du+\int_t^s\eta'(v^{t,v}_u)dB^1_u\Big).
$$
We finally recall that,  under \textbf{(H)}, all above processes $\xi^{t,x,v},v^{t,v},Y,Z$ can be collected in a multidimensional diffusion process $X^{t,x,v}$, whose diffusion coefficient and drift are either Lipschitz continuous and with sublinear growth. Thus, in particular, standard $L^p$ estimates hold, as the Burkholder one: for every $p\geq 1$, $t\in [0,T]$ and $x,v\in\R$ there exists a positive constant  $C$ such that
$$
\E(\sup_{t\leq s\leq T}|X_s^{t,x,v}|^p)\leq C\,(T-t)^{p/2}(1+|x|^p+|v|^p).
$$

\section{Antonelli and Scarlatti power series expansion}
\label{sect-as}

We recall here the result obtained by Antonelli and Scarlatti in \cite{ScarlattiAntonelli}. They proved a development in power series of the European call option price-function with respect to the correlation between the noise of the risky asset and the noise associated to the stochastic volatility.

\smallskip

Consider a European call option written on our market model. For a fixed maturity $T$ and strike price $K$, its price-function is given by
\begin{equation}\label{def-u}
u(t,x,v;\rho)=\mathbb{E}\left(e^{-r(T-t)}\left(e^{\xi_T^{t,x,v}}-K\right)_+\right).
\end{equation}
When the underlying asset price process evolves following the Black and Scholes \cite{BS} model,
(\ref{def-u}) has a closed form solution, following  the worldwide famous formula
\begin{equation}\label{BS-formula}
BS(t,x,\sigma)=e^xN(d_1(t,x,\sigma))-Ke^{-r{(T-t)}}N(d_2(t,x,\sigma)),
\end{equation}
where
\begin{equation}
\label{defd2}
d_2(t,x,\sigma)=\frac{x-\ln K+(r-\sigma^2/2)(T-t)}{\sigma\sqrt{T-t}}
\quad\mbox{and}\quad d_1=d_2+\sigma\sqrt{T-t}.
\end{equation}

In order to tackle the problem of the regularity of the price-function $u$ given in (\ref{def-u}) w.r.t. $\rho$,  the following assumption is considered in \cite{ScarlattiAntonelli}:
\begin{itemize}
\item[\textbf{(AS)}] \emph{For every $t\in[0,T)$ and $v\in\R$ there exists a constant $C_1>0$ such that for any $q\ge1$,}
$$
\mathbb{E}\Big(\Big(\int_t^T|f(v_r^{t,v})|^2dr\Big)^{-q}\Big)\le C_1^q.
$$
\end{itemize}
Then, the following result is stated in Theorem 2.2 of Antonelli and Scarlatti \cite{ScarlattiAntonelli}:
\begin{thm} \label{teoConv} Assume that \textbf{(H)} and \textbf{(AS)} both hold. Let $u=u(t,x,v;\rho)$ denote the price-function (\ref{def-u}). Then,
\begin{itemize}
\item[(i)]
the map $\rho \mapsto u(t,x,v;\rho)$ is infinitely differentiable in a neighborhood of $\rho=0$;
\item[(ii)]
there exists $R\in(0,1)$ such that for $|\rho|<R$ one has
$$
u(t,x,v;\rho)=\sum\limits_{k\ge 0}\frac 1 {k!}\partial^k_\rho u(t,x,v;\rho)_{|\rho=0}\,  \rho^k.
$$
\end{itemize}
\end{thm}

For $k\geq 0$, let
$$
g_k(t,x,v):=\frac 1{k!}\partial^k_\rho u(t,x,v;\rho){|_{\rho=0}}
$$
denote the coefficients of the Taylor expansion of $u$. In \cite{ScarlattiAntonelli},
it is shown that the $g_k$'s solve suitable parabolic partial differential equations and, by using Feynman-Kac type formulas, each $g_k$ can be represented in terms of an expectation of a suitable functional of the diffusion process $(\hat \xi,v)$ (see (3.1) in \cite{ScarlattiAntonelli}). Moreover, by conditioning w.r.t. the noise $B^1$, such expectations reduce to ones of functionals of the volatility process $v$ only. Here, we write down the formulas for $g_k$ as $k=0,1,2$ (for the general formulation, see
(3.3) in \cite{ScarlattiAntonelli}): one has
\begin{align}
g_0(t,x,v)
=&\mathbb{E}\left(BS\bigg(t,x,\sqrt{\frac{\<M^{t,v}\>_{[t,T]}}{T-t}}\bigg)\right)\label{gg0}\\
g_1(t,x,v)
=&-Ke^{-r(T-t)}\mathbb{E}\Bigg([d_2N'(d_2)]\bigg(t,x,\sqrt{\frac{\<M^{t,v}\>_{[t,T]}}{T-t}}\bigg)\frac {c_{[t,T]}^{t,v}}{\<M^{t,v}\>_T}\Bigg)\label{gg1}\\
g_2(t,x,v)
=&Ke^{-r(T-t)}\int_t^T\mathbb{E}\Bigg(\frac {(f\eta)(v_{\alpha_2}^{t,v})}{\big(\<M^{t,v}\>_{[t,T]}\big)^{\frac 32}}\Big(\partial_v v_{\alpha_2}^{t,v}\Big)^{-1}\times \label{gg2}\\
&\times\Bigg[N^{(3)}\bigg(d_2\bigg(t,x,\sqrt{\frac{\<M^{t,v}\>_{[t,T]}}{T-t}}\bigg)\bigg)
\partial_v c_{[\alpha_2,T]}^{t,v}+\frac{c_{[\alpha_2,T]}^{t,v}}{2}
\partial_v\<M^{t,v}\>_{[\alpha_2,T]}\times\nonumber\\
&\times\Bigg(\frac{N^{(5)}(d_2)}{\<M^{t,v}\>_{[t,T]}}-\frac{N^{(4)}(d_2)}{\sqrt{\<M^{t,v}\>_{[t,T]}}}\Bigg)\bigg(t,x,\sqrt{\frac{\<M^{t,v}\>_{[t,T]}}{T-t}}\bigg)\Bigg]\Bigg)d\alpha_2\nonumber
\end{align}
where $N$ denotes the cumulative distribution function associated with the standard Gaussian law and $N^{(n)}(\cdot)$ is its $n$th derivative, $d_2$ is given in (\ref{defd2}) and for $ s\in[t,T]$,
\begin{align}
\label{defcuT}
c_{[s,T]}^{t,v}&=\int_s^T(f\eta)(v_{\alpha}^{t,v})\int_{\alpha}^T
(ff')(v_\beta^{t,v})\partial_v v_\beta^{t,v}\left(\partial_v v_{\alpha}^{t,v}\right)^{-1}\!\!\!\!\,d\beta d\alpha,\\
\label{defM2}
\<M^{t,v}\>_{[s,T]}&=\int_s^T \big(f(v_\alpha^{t,v})\big)^2d\alpha.
\end{align}

\section{Series expansion A - from the Al\acc os formula}\label{sect-alos}

We deal here with the price-formula obtained by Al\`os in \cite{Alos1} and we derivate it in order to state representation formulas for the series expansion.

\smallskip

Consider a European option with maturity $T>0$ and payoff $h(\xi_T)$, whose price, as seen at time $t<T$, is given by
\begin{equation}\label{price-h}
V_t=\mathbb{E}\left(e^{-r(T-t)}h(\xi_T)|\mathcal{F}_t\right).
\end{equation}
We denote through $BS_h(t,x;\sigma)$ the price-function of this option under the Black and Scholes model with constant volatility $\sigma$, that is
\begin{equation}\label{priceBS-h}
BS_h(t,x;\sigma)=\mathbb{E}\Big( e^{-r(T-t)}h\Big(e^{x+(r-\frac{\sigma^2}{2})(T-t)+\sigma (W_T-W_t)}\Big)\Big),
\end{equation}
$W$ denoting a standard Brownian motion on $\R$.

\smallskip

In order to state the starting result from Al\`os \cite{Alos1}, one needs some requirements (see (H1)-(H4) therein, at page 357). In our context, which takes into account Assumption \textbf{(H)}, we can rewrite them as follows:
\begin{itemize}
\item[{\bf(P)}]
\emph{the payoff function $h:\mathbb{R}\rightarrow \mathbb{R}_+$ is continuous and $\mathcal{C}_b^1$-piecewise;}
\item[{\bf(A)}]
\emph{there exists $a>0$ such that $|f|^2>a$.}
\end{itemize}
Let us remark that in Al\`os \cite{Alos1} the model is more general (the volatility is not necessary a diffusion process, neither Markovian, but definitively measurable w.r.t. the noise $B^1$). Moreover, a further request appears (see hypotheses (H4) therein, at page 357), which is not written in above \textbf{(P)} and \textbf{(A)}  because it can be actually dropped (see e.g. the more general result - jumps are allowed - in Al\`os, Le\'on and Vives \cite{Alos3}, Theorem 4.2 therein).

Theorem 3 in Al\`os \cite{Alos1} states the following generalization of the Hull and White formula \cite{HullandWhite}.

\begin{thm}\label{alos-thm}
Assume that \textbf{(H)}, \textbf{(P)} and \textbf{(A)} hold. Let $V$ and $BS_h$ be given by (\ref{price-h}) and (\ref{priceBS-h}) respectively.
Then, for any $t\in [0,T]$, one has
\begin{equation} \label{m}
V_t=\mathbb{E}\Bigg( BS_h\bigg(t,\xi_t;\sqrt{\frac{\<M\>_{[t,T]}}{T-t}}\bigg)\,\Big\vert\,\mathcal{F}_t\Bigg)
+\frac{\rho}{2}\mathbb{E}\Bigg(\int_t^Te^{-r(s-t)}H\bigg(s,\xi_s;\sqrt{\frac{\< M \>_{[s,T]}}{T-s}}
\bigg)\Lambda_sds\,\Big\vert\,\mathcal{F}_t\Bigg)
\end{equation}
where, as $s\in [t,T]$,
\begin{align}
\< M \>_{[s,T]}&=\int_s^T \big(f(v_r)\big)^2\,dr\label{MsT}\\
H(s,x;\sigma)&=\big(\partial_x^3-\partial_x^2\big)BS_h(s,x;\sigma)\label{H}\\
\Lambda_s&=\left(\int_s^T D^1_s\big((f(v_r))^2\big)\,dr\right)f(v_s) \label{lambda}
\end{align}
\end{thm}

The notation $D^1$ in (\ref{lambda}) stands for the Malliavin derivative in the direction of $B^1$. We do not enter here in the details of Malliavin calculus (we will do this in Section \ref{sect-mall}).

\smallskip

Let us now specify to our problem. First, we notice that \textbf{(A)} implies that \textbf{(AS)} holds. And obviously, \textbf{(P)} is fulfilled by the call option payoff.
Then, we can proceed to study the Taylor series expansion of the call price-function $u$ by developing formula (\ref{m})  w.r.t. $\rho$. To this purpose, let us recall the notation $BS$ for the call option price-function given by the Black and Scholes formula, see (\ref{BS-formula}).
We first state the following result.
\begin{prop}\label{prop-alos}
Assume that \textbf{(H)} and \textbf{(A)} hold. Then the European call option price-function $u$ given in (\ref{def-u}) can be written as
\beq\label{alfa_beta}
u(t,x,v;\rho)=\mathbb{E}\big(\Gamma^{t,x,v}_1\big)
+\rho\mathbb{E}(\Gamma^{t,x,v}_2(\rho))
\eeq
where $\Gamma^{t,x,v}_1$ and $\Gamma^{t,x,v}_2(\rho)$ are defined as
\begin{align}
\label{Gamma1}
\Gamma^{t,x,v}_1
&=
BS\bigg(t,x,\sqrt{\frac{\<M^{t,v}\>_{[t,T]}}{T-t}}\bigg)\\
\label{Gamm2tilde}
\Gamma^{t,x,v}_2(\rho)
&= - K e^{-r(T-t)}\int_t^T \frac{\psi^{t,v}_s}{\<M^{t,v}\>_{[t,T]} - \rho^2\<M^{t,v}\>_{[t,s]}}[D^{t,x,v}N^{'}(D^{t,x,v})](s;\rho) ds
\end{align}
$\<M^{t,v}\>_{[s,T]}$, $s\in [t,T]$, being given in (\ref{defM2}) and for $s\in [t,T]$,
\begin{align}
\label{psi}
\psi_s^{t,v}
&=(f\eta) (v_s^{t,v})\int_s^T(ff')(v_r^{t,v})\partial_v v_r^{t,v}\left(\partial_v v_s^{t,v}\right)^{-1}\! dr,\\
\label{D}
D^{t,x,v}(s;\rho)
&=\frac{\<M^{t,v}\>^{1/2}_{[t,T]}\,d_2\Big(t,x,\sqrt{\frac{\<M^{t,v}\>_{[t,T]}}{T-t}}\Big) + \rho\int_t^sf(v^{t,v}_u) dB^{1}_u}{\big(\<M^{t,v}\>_{[t,T]} - \rho^2\<M^{t,v}\>_{[t,s]}\big)^{1/2}}
\end{align}
\end{prop}


The proof of  Proposition \ref{prop-alos} is an immediate consequence of the following result.

\begin{lemma}\label{prop-alos-appendix1}
Under assumptions \textbf{(H)} and \textbf{(A)}  the following statements hold:
\begin{itemize}
\item[$(i)$]
the European call option price-function $u$ given in (\ref{def-u}) can be written as
\beq\label{alfa_beta-bis}
u(t,x,v;\rho)=\mathbb{E}\big(\Gamma^{t,x,v}_1\big)
+\rho\mathbb{E}(\bar\Gamma^{t,x,v}_2(\rho))
\eeq
where $\Gamma^{t,x,v}_1$ is given in (\ref{Gamma1}) and
\begin{align}
\label{Gamma2}
\bar\Gamma^{t,x,v}_2(\rho)
&=-Ke^{-r(T-t)}\int_t^T\frac1{\<M^{t,v}\>_{[s,T]}}\, [d_2N'(d_2)]\bigg(s,\xi_s^{t,x,v};\sqrt{\frac{\<M^{t,v}\>_{[s,T]}}{T-s}}\bigg)
\,\psi_s^{t,v}\,ds
\end{align}
$\<M^{t,v}\>_{[s,T]}$  and $\psi_s^{t,v}$ being given in (\ref{defM2}) and (\ref{psi}) respectively;
\item[$(ii)$]
if $\bar\Gamma_2^{t,x,v}$ and $\Gamma_2^{t,x,v}$ are as in (\ref{Gamma2}) and (\ref{Gamm2tilde}) respectively, one has
\beq
\label{EGamma2tilde}
\mathbb{E}(\bar\Gamma^{t,x,v}_2(\rho)) = \mathbb{E}(\Gamma^{t,x,v}_2(\rho))
\eeq
\end{itemize}
\end{lemma}

\emph{Proof. }
\textbf{$(i)$}
As already observed, requirements \textbf{(H)} and \textbf{(A)} imply that both Theorem \ref{teoConv} and Theorem \ref{alos-thm} hold. So we can actually proceed with the proof.
For a European call option, the price-function $u$ is given through $u(t,\xi_t,v_t;\rho)=V_t$, where $V_t$ is as in formula (\ref{m}) with $h(\xi)=(e^\xi-K)_+$.
We firstly notice that from (\ref{MsT}) and (\ref{defM2}) one has $\<M \>_{[s,T]}=\<M^{t,v_t }\>_{[s,T]}$. So, by using the Markov property one immediately gets
$$
\mathbb{E}\Bigg( BS\bigg(t,\xi_t;\sqrt{\frac{\< M \>_{[t,T]}}{T-t}}\bigg)\bigg|
\mathcal{F}_t\Bigg)
= \mathbb{E}\Bigg( BS\bigg(t,x,\sqrt{\frac{\<M^{t,v}\>_{[t,T]}}{T-t}}\bigg)\Bigg)\bigg|_{\mbox{\scriptsize{$\begin{array}{l}x=\xi_t\\v=v_t\end{array}$}}}.
$$
Therefore, we prove that  for any $s\in (t,T]$,
\begin{equation}\label{tobeproved2}
\begin{array}{l}
\displaystyle
\mathbb{E}\Bigg(e^{-r(s-t)}H\bigg(s,\xi_s;\sqrt{\frac{\< M \>_{[s,T]}}{T-s}}\bigg)\Lambda_s\Big\vert\mathcal{F}_t\Bigg)
= \smallskip\\
\quad=-2\displaystyle
\mathbb{E}\Bigg(Ke^{-r(T-t)}\frac1{\<M^{t,v}\>_{[s,T]}}
[d_2N'(d_2)]\bigg(s,\xi_s^{t,x,v};\sqrt{\frac{\<M^{t,v}\>_{[s,T]}}{T-s}}\bigg)\psi_s^{t,v}
\Bigg)\Bigg|_{\mbox{\scriptsize{$\begin{array}{l}x=\xi_t\\v=v_t\end{array}$}}}
\end{array}
\end{equation}
where $H$ and $\Lambda$ are given by (\ref{H}) (with $BS_h$ replaced by $BS$) and (\ref{lambda}) respectively. And moreover,  we prove also that one can actually use the Fubini theorem, so that (\ref{Gamma2}) will hold.

\smallskip

By recalling (\ref{BS-formula}) and by using the well known properties $\partial_xBS(t,x,\sigma)=e^xN(d_1)$, $N''(\tau)=-\tau N'(\tau)$ and
$N'(d_1)=N'(d_2)K e^{-x-r(T-t)}$,
straightforward computations give
$$
\big(\partial^3_x-\partial^2_x\big)BS(t,x,\sigma)=
-K\frac{e^{-r(T-t)}}{\sigma^2(T-t)}\cdot[d_2 N'(d_2)](t,x,\sigma)
$$
so that
$$
H\bigg(s,\xi_s;\sqrt{\frac{\<M \>_{[s,T]}}{T-s}}\bigg)=-K\frac{e^{-r(T-s)}}{\<M \>_{[s,T]}}\cdot[d_2N'(d_2)]\bigg(s,\xi_s;\sqrt{\frac{\<M \>_{[s,T]}}{T-s}}\bigg).
$$
Let us consider $\Lambda$ as in (\ref{lambda}). First, thanks to \textbf{(H)}, for $r>u\geq t$ one has
$$
D^1_s f^2(v_r)
=2(ff')(v_r)D^1_s v_r\quad\mbox{and}\quad
D^1_s v_r
=\Big(\partial_v v_r^{t,v}
\big(\partial_v v_s^{t,v}\big)^{-1}\Big)\Big|_{v=v_t}
\eta(v_s),
$$
$\partial_v v_s^{t,v}$, $s\geq t$, denoting the first variation process of $v^{t,v}$, see (\ref{dv}).
Therefore,
$$
\Lambda_s
=(f\eta)(v_s)\int_s^T2(ff')(v_r) \Big(\partial_v v_r^{t,v}
\big(\partial_v v_s^{t,v}\big)^{-1}\Big)\Big|_{v=v_t}dr
=2\psi_s^{t,v_t},
$$
the process $\psi^{t,v}$ being given by (\ref{psi}). By inserting the expressions for
$H(s,\xi_s;\sqrt{\frac{\<M \>_{[s,T]}}{T-s}})$ and $\Lambda_s$
we get
$$
\begin{array}{l}
\displaystyle
\hskip -0.5cm
e^{-r(s-t)}H\bigg(s,\xi_s;\sqrt{\frac{\<\bar M \>_{[s,T]}}{T-s}}\bigg)\Lambda_s=\smallskip\\
\displaystyle
=-2K\frac{e^{-r(T-t)}}{\<M^{t,v_t}\>_{[s,T]}}
\cdot[d_2N'(d_2)]\bigg(s,\xi_s^{t,\xi_t,v_t};\sqrt{\frac{\<M^{t,v_t}\>_{[s,T]}}{T-s}}\bigg)
\,\psi^{t,v_t}_s
=: Z_s
\end{array}
$$
and the Markov property allows to conclude that (\ref{tobeproved2}) really holds. It remains to prove that the Fubini theorem can be applied. So, we show now that for every $p\geq 1$, the above process $Z$ belong to $L^p$ w.r.t. the product measure $d\P\times ds$.

Since $|xN'(x)| = |N^{(2)}(x)|$ and $\sup_x|xN'(x)|\leq const$ (from now on, $const$ denotes a suitable constant, possibly varying) and since \textbf{(A)} gives $\<M^{t,v}\>_{[s,T]}\geq a(T-s)$, one has
$$
\int_t^T|Z_s|^p
\leq const\cdot \int_t^T\frac1{(T-s)^{p}}\,
\,|\psi_s^{t,v}|^p\,ds.
$$
Then, it remains to prove that $\E(\int_t^T\frac1{(T-s)^{p}}\,
\,|\psi_s^{t,v}|^p\,ds)<\infty$. Notice that in particular this gives that $\psi^{t,v}\in L^p$ w.r.t. $d\P\times ds$.
One has
$$
|\psi_s^{t,v}|^p\leq
const\cdot  (1+|v_s^{t,v}|^p) \left|\partial_v v_s^{t,v}\right|^{-p}\,(T-s)^{p-1}\,
\int_s^T(1+|v_r^{t,v}|^p)|\partial_v v_r^{t,v}|^p\, dr
$$
so that
$$
\int_t^T\frac1{(T-s)^{p}}\,
\,|\psi_s^{t,v}|^p\,ds\leq
\int_t^T\frac{\int_s^T(1+|v_r^{t,v}|^p)|\partial_v v_r^{t,v}|^p\, dr}{(T-s)}\,
\,(1+|v_s^{t,v}|^p) |\partial_v v_s^{t,v}|^{-p}\,ds.
$$
By using the Burkholder inequality, we know that $\sup_{t\leq s\leq T}(1+|v_s^{t,v}|^p) |\partial_v v_s^{t,v}|^{-p}\in L^q$ for all $q$. So, by using the H\"older inequality, it remains to prove that for $\alpha>1$,
$$
\int_t^T\frac{1}{(T-s)}\,
\E\Big(\Big(\int_s^T(1+|v_r^{t,v}|^p)|\partial_v v_r^{t,v}|^p\, dr\Big)^\alpha\Big)^{1/\alpha} <\infty.
$$
By the Jensen inequality one gets
\begin{align*}
\E\Big(\Big(\int_s^T(1+|v_r^{t,v}|^p)|\partial_v v_r^{t,v}|^p\, dr\Big)^\alpha\Big)^{1/\alpha}
\leq & (T-s)^{1-\frac 1\alpha}
\E\Big(\int_s^T(1+|v_r^{t,v}|^p)|^\alpha\partial_v v_r^{t,v}|^{p\alpha}\, dr\Big)^{1/\alpha}\\
\leq &
(T-s)\E\Big(\sup_{t\leq r\leq T}
(1+|v_r^{t,v}|^p)^\alpha\,|\partial_v v_r^{t,v}|^{p\alpha}\Big)^{1/\alpha}\\
\leq & const\times (T-s)
\end{align*}
and this gives the result.
Statement $(i)$ is now completely proved.

\smallskip

\textbf{$(ii)$}
By applying the Fubini theorem, we have
\beq
\label{Gamm2inter}
\begin{array}{ll}
&\mathbb{E}(\bar\Gamma^{t,x,v}_2(\rho))=-Ke^{-r(T-t)}\times\smallskip\\
&\displaystyle
\hskip 1cm \times\int_t^T
\mathbb{E}\Big(\frac{\psi_s^{t,v}}{\<M^{t,v}\>_{[s,T]}}\,\mathbb{E}\Big( [d_2N'(d_2)]\bigg(s,\xi_s^{t,x,v};\sqrt{\frac{\<M^{t,v}\>_{[s,T]}}{T-s}}\bigg)\,\Big\vert\,\mathcal{F}^{B^1}_T
\Big)\Big)\,ds
\end{array}
\eeq
where ${\cal F}^{B^1}_T=\sigma(B^1_s\,:\,s\leq T)$. We notice that
\begin{align*}
&d_2\Big(s,\xi_s^{t,x,v},\sqrt{\frac{\<M^{t,v}\>_{[s,T]}}{T-s}}\Big)=
\nonumber \\
&=\frac{\sqrt{\<M^{t,v}\>_{[t,T]}}d_2\Big(t,x,\sqrt{\frac{\<M^{t,v}\>_{[t,T]}}{T-t}}\Big) + \rho \int_t^s f(v^{t,v}_{\theta})dB^1_{\theta}}{\sqrt{\<M^{t,v}\>_{[s,T]}}} + \sqrt{1 - \rho^2}\frac{\int_t^s f(v^{t,v}_{\theta})dB^2_{\theta}}{\sqrt{\<M^{t,v}\>_{[s,T]}}},
\end{align*}
Hence, conditionally on $\mathcal{F}^{B^1}_T$, $d_2\Big(s,\xi_s^{t,x,v},\sqrt{\frac{\<M^{t,v}\>_{[s,T]}}{T-s}}\Big)$ is a Gaussian r.v. with variance and mean given by
$$
\begin{array}{l}
\displaystyle
{\sigma^{t,v}(s;\rho)}^2=(1 - \rho^2)\frac{\<M^{t,v}\>_{[t,s]}}{\<M^{t,v}\>_{[s,T]}}\smallskip\\
\displaystyle
\mu^{t,x,v}(s;\rho)=\frac{\sqrt{\<M^{t,v}\>_{[t,T]}}
d_2\Big(t,x,\sqrt{\frac{\<M^{t,v}\>_{[t,T]}}{T-t}}\Big) + \rho \int_t^s f(v^{t,v}_{\theta})dB^1_{\theta}}{\sqrt{\<M^{t,v}\>_{[s,T]}}}.
\end{array}
$$
respectively. Now, if $Z\sim\mathcal{N}( \mu,\sigma^2)$ , straightforward computations
give
$$
\E\Big(Ze^{-Z^2/2}\Big)
=\frac{ \mu }{(1 + \sigma^2)^{\frac{3}{2}}}e^{-\frac{\mu^2} {2(1+\sigma^2)}}.
$$
Thus, by inserting $\sigma=\sigma^{t,v}(s;\rho)$ and $\mu=\mu^{t,v}(s;\rho)$,
one gets
\beq
\label{d2cond}
\begin{array}{l}
\displaystyle
\mathbb{E}\Big( [d_2N'(d_2)]\Big(s,\xi_s^{t,x,v};\sqrt{\frac{\<M^{t,v}\>_{[s,T]}}{T-s}}\Big)
\,\Big\vert\,\mathcal{F}^{B^1}_T\Big)=\smallskip\\
\displaystyle
\hskip 2cm=\frac{\<M^{t,v}\>_{[s,T]}}{\<M^{t,v}\>_{[t,T]}-\rho^2\<M^{t,v}\>_{[t,s]}}
[D^{t,x,v}N^{'}(D^{t,x,v})](s;\rho),
\end{array}
\eeq
where $D^{t,x,v}(s;\rho)$ is defined in (\ref{D}).
%
By putting (\ref{d2cond}) in (\ref{Gamm2inter}), the use of the Fubini theorem allows to get the claimed result. So, in order to conclude we must only to show that the Fubini theorem can be actually used, and this holds if
$$
\mathbb{E}\left(\int_t^T\,\Big\vert\, K e^{-r(T-t)}\frac{\psi^{t,v}_s}{\<M^{t,v}\>_{[t,T]} - \rho^2\<M^{t,v}\>_{[t,s]}}[D^{t,x,v}N^{'}(D^{t,x,v})](s;\rho)\,\Big\vert ds\right) < + \infty.
$$
But this is true because $\frac{1}{\<M^{t,v}\>_{[t,T]} - \rho^2\<M^{t,v}\>_{[t,s]}}[D^{t,x,v}N^{'}(D^{t,x,v})](s;\rho)$ is bounded and $\psi^{t,v}\in L^p$ w.r.t. $d\P\times ds$.
\cvd

\smallskip

Now, under \textbf{(H)} and \textbf{(A)}, Theorem \ref{teoConv} can be applied, so that in particular $\rho\mapsto\mathbb{E}(\Gamma^{t,x,v}_2(\rho))$ is actually $C^\infty$ in a neighborhood of $\rho=0$ (we stress that $\Gamma^{t,x,v}_1$ is independent of $\rho$).
So, in order to give an alternative representation of the coefficients of the Taylor expansion, we compute the coefficients
\begin{equation}\label{uk}
u_k(t,x,v)=\partial_\rho^k u(t,x,v;\rho)_{\big|\rho=0}
=\partial_\rho^k\Big(\rho\mathbb{E}(\Gamma^{t,x,v}_2(\rho))
\Big)_{\big|\rho=0},\quad k\geq 1
\end{equation}
by taking the derivatives inside the expectations, and this can be done.
Of course, $u_k=k!g_k$, the $g_k$'s as in Theorem \ref{teoConv}, althought their representations in terms of expectations are not generally the same. In fact, we have

\begin{prop}\label{derivataGamma2tilde}
Assume that \textbf{(H)} and \textbf{(A)} hold. Let $\Gamma^{t,x,v}_2(\rho)$ be defined in (\ref{Gamm2tilde}). Then for every $k\geq 0$ one has
$$
\partial^k_\rho
\E\big(\Gamma^{t,x,v}_2(\rho)\big)_{|_{\rho=0}}
= K e^{-r(T-t)}
\E\Big(\frac 1{\<M^{t,v}\>_{[t,T]}}\int_t^T \Xi^{t,x,v}_k(s)\,\psi_s^{t,v}\,ds\Big)
$$
with
$$
\Xi^{t,x,v}_k(s)
=k!\sum_{j=0}^k
G_{1,k-j}^{t,v}(s)\,
G_{2,j}^{t,x,v}(s)
$$
in which $\{G^{t,x,v}_{1,k}\}_k$ and $\{G^{t,x,v}_{1,k}\}_k$ are given by
\begin{equation}
\label{G1n}
G^{t,v}_{1,\ell}(s)=\I_{\{\ell \mbox{\scriptsize{ even}}\}}\,
\Big(\frac{\<M^{t,v}\>_{[t,s]}}{\<M^{t,v}\>_{[t,T]}}\Big)^{\ell/2},\quad
\ell=0,1,\ldots
\end{equation}
and
\beq
\label{G2n}
\begin{array}{ll}
\ell=0:\quad&
G^{t,x,v}_{2,0}(s;\rho)=
[N^{(2)}(d_2)] \Big(t,x,\sqrt{\frac{\<M^{t,v}\>_{[t,T]}}{T-t}}\Big)\smallskip\\
\ell\geq 1:\quad&
\displaystyle
G^{t,x,v}_{2,\ell}(s;\rho)= \sum_{\nu=1}^{\ell}
\frac{1}{\nu!}\,
[N^{(2+\nu)}(d_2)] \Big(t,x,\sqrt{\frac{\<M^{t,v}\>_{[t,T]}}{T-t}}\Big)\!\!\!\!\!\!\!\!\!\!\!\sum_{\mbox{\scriptsize $ \begin{array}{c}h_i\geq 1\ \forall i\\ h_1 + ...+ h_{\nu}=\ell\end{array}$}}\!\!\!\!
\prod_{i=1}^{\nu} D^{t,x,v}_{h_i}(s)
\end{array}
\eeq
respectively, where
\begin{equation}\label{Dn}
D^{t,x,v}_{h}(s)=\left\{
\begin{array}{ll}
\displaystyle
\frac{(2n)!}{(n!)^2\,4^n}
\Big(\frac{\<M^{t,v}\>_{[t,s]}}{\<M^{t,v}\>_{[t,T]}}\Big)^n
d_2\Big(t,x,\sqrt{\frac{\<M^{t,v}\>_{[t,T]}}{T-t}}\Big)
&\mbox{ if } h=2n,\smallskip\\
\displaystyle
\frac{(2n)!}{(n!)^2\,4^n}
\Big(\frac{\<M^{t,v}\>_{[t,s]}}{\<M^{t,v}\>_{[t,T]}}\Big)^n
\,\frac{\int_t^sf(v^{t,v}_u)\,dB^1_u}{\<M^{t,v}\>^{1/2}_{[t,T]}}
&\mbox{ if } h=2n+1.
\end{array}
\right.
\end{equation}
As a consequence, one has
$$
u_n(t,x,v):=\partial^n_\rho u(t,x,v;\rho)_{|_{\rho=0}}
=n K e^{-r(T-t)}
\E\Big(\frac 1{\<M^{t,v}\>_{[t,T]}}\int_t^T \Xi^{t,x,v}_{n-1}(s)\,\psi_s^{t,v}\,ds\Big)
,\quad n\geq 1.
$$
\end{prop}

\proof
One has
$$
\Gamma^{t,x,v}_2(\rho)
= \frac{K e^{-r(T-t)}}{\<M^{t,v}\>_{[t,T]}}\int_t^T G_1^{t,x,v}(s;\rho)\,G_2^{t,x,v}(s;\rho)\,\psi_s^{t,v}\,ds
$$
with
$$
G_1^{t,v}(s;\rho)=
\frac{\<M^{t,v}\>_{[t,T]}}{\<M^{t,v}\>_{[t,T]} - \rho^2\<M^{t,v}\>_{[t,s]}}
\quad\mbox{and}\quad
G_2^{t,x,v}(s;\rho)=N''(D^{t,x,v})(s;\rho)
$$
(recall that $-\tau N'(\tau)=N''(\tau)$).
We write formally
\begin{align*}
\partial^k_\rho
\E\big(\Gamma_2^{t,x,v}(\rho)\big)
&= K e^{-r(T-t)}\sum_{j=0}^k\Big(\begin{array}{c}k\\j\end{array}\Big)\E\Big(
\frac 1{\<M^{t,v}\>_{[t,T]}}
\int_t^T \partial^{k-j}_\rho
G_1^{t,x,v}(s;\rho)\,\partial^{j}_\rho
G_2^{t,x,v}(s;\rho)\psi_s^{t,v}\,ds\Big)
\end{align*}
So, in order to prove our result it is sufficient that if $\rho$ is close to 0 then
$$
\E\Big(
\int_t^T \big|\partial^{k-j}_\rho
G_1^{t,x,v}(s;\rho)\,\partial^{j}_\rho
G_2^{t,x,v}(s;\rho)\psi_s^{t,v}\big|\,ds\Big)<\infty
$$
(recall that \textbf{(A)} gives $1/{\<M^{t,v}\>_{[t,T]}}\leq const$).
In the proof of Lemma \ref{prop-alos-appendix1} we have already proved that
$\E(
\int_t^T \big|\psi_s^{t,v}|^p\,ds)<\infty$ for all $p$. Therefore, it remains
to show that there exists $R>0$ such that if $|\rho|<R$ then
$$
\E\Big(
\int_t^T \big|\partial^{k-j}_\rho
G_1^{t,x,v}(s;\rho)\big|^p\,ds\Big)<\infty
\quad\mbox{and}\quad
\E\Big(
\int_t^T \big|\partial^{j}_\rho
G_2^{t,x,v}(s;\rho)\big|^p\,ds\Big)<\infty
$$
for all  $p$. So, we prove these facts.
And to simplify notations, we drop the dependence on $t,x,v$.

\smallskip

From $(1-t)^{-1}=\sum_{n\geq 0}t^n$, $|t|<1$, one immediately gets
$$
G_1(s;\rho)=\sum_{k\geq 0}G_{1,k}(s)\,\rho^k
\quad \mbox{where $\{G_{1,k}(s)\}_k$ is given in (\ref{G1n}).}
$$

Since $\sup_k\sup_{s\in[t,T]}|G_{1,k}(s)|\leq 1$, we can write
$
\partial^\ell_\rho
G_1(s;\rho)=\sum_{k\geq 0}G_{1,k}(s)\,\partial^\ell_\rho \rho^k.
$
So, for any $0<R<1$ and $|\rho|\leq R$,
$$
|\partial^\ell_\rho
G_1(s;\rho)|\leq \sum_{k\geq \ell}\,\frac{k!}{(k-\ell)!}|\rho|^{k-\ell}
= \ell!\,(1-|\rho|)^{-\ell-1}\leq C_{\ell,R}
$$
$C_{\ell,R}$ being a suitable positive constant, and this gives the first result. Let us now study $G_2(s;\rho)$. We can write
\beq
\label{G2deriv}
\partial_{\rho}^{j}G_2(s;\rho)=\partial_{\rho}^{j}[N''(D)](s;\rho)
=j!\sum_{\nu=1}^{j}
\frac{N^{(2+\nu)}(D(s;\rho))}{\nu!}\!\!\!\!\!\!\!\!\!\!\!\sum_{\mbox{\scriptsize $ \begin{array}{c}h_i\geq 1\ \forall i\\ h_1 + ...+ h_{\nu}=j\end{array}$}}\!\!\!\!\!\!\!
\prod_{i=1}^{\nu}\frac{1}{h_i !}
\partial_\rho^{h_i}D(s;\rho).
\eeq
Now, for $k\geq 0$ one has $N^{(1+k)}(\tau)=(-1)^k H_k(\tau)N'(\tau)$
where $H_k$ denotes the Hermite polynomial of order $k$. So, for some positive constant $L_j$, it holds that $\sup_x|N^{(2+\nu)}(x)|\leq L_j$ for every $\nu=1,\ldots,j$.
Therefore,
\begin{equation}\label{app1}
|\partial_{\rho}^{j}G_2(s;\rho)|\leq
L_j j!\sum_{\nu=1}^{j}
\frac{1}{\nu!}\!\!\!\!\!\sum_{\mbox{\scriptsize $ \begin{array}{c}h_i\geq 1\ \forall i\\ h_1 + ...+ h_{\nu}=j\end{array}$}}\!\!\!\!\!\!\!
\prod_{i=1}^{\nu}\frac{1}{h_i !}
\big|\partial_\rho^{h_i}D(s;\rho)\big|.
\end{equation}

Consider now $D(s;\rho)$.
Since for $|t|<1$ one has $(1-t^2)^{-1/2}=\sum_{n\geq 0}c_n\,t^{n}$, with $c_n=\frac{(2n)!}{(n!)^2\,4^n}$, by inserting in formula (\ref{D}) for $|\rho|<1$ one gets
$$
D(s;\rho)
=\sum_{k\geq 0}D_k(s)\,\rho^k
\quad \mbox{where $\{D_{k}(s)\}_k$ is given in (\ref{Dn}).}
$$
And since for every $0<R<1$ there exists $C>0$ such that $|c_n|\leq C R^{-n}$, we obtain
$$
|D_{k}(s)|\leq C\,M_s\,R^{-k},\quad
\mbox{with}\quad
M_s=
\Big|d_2\Big(t,x,\sqrt{\frac{\<M\>_{[t,T]}}{T-t}}\Big)
\Big|+\frac{\Big|\int_t^sf(v_u)\,dB^1_u\Big|}{\<M\>_{[t,T]}}
$$
Moreover, for $|\rho|<1$ one has
$\partial^\ell_\rho D(s;\rho)
=\sum_{k\geq 0}D_k(s)\,\partial^\ell_\rho\rho^k$. So,
if we take $|\rho|<R/2$ we can write
$$
|\partial^\ell_\rho D(s;\rho)|
\leq \sum_{k\geq \ell} C\,M_s\,R^{-k}\,\frac{k!}{(k-\ell)!}|\rho|^{k-\ell}
\leq  \ell ! \,2^{\ell+1} C M_s R^{-\ell}.
$$

As a consequence, we get
$$
|\partial_{\rho}^{j}G_2(s;\rho)|\leq
j! L_j \sum_{\nu=1}^{j}
\frac{1}{\nu!}\!\!\!\!\!\sum_{\mbox{\scriptsize $ \begin{array}{c}h_i\geq 1\ \forall i\\ h_1 + ...+ h_{\nu}=j\end{array}$}}\!\!\!\!\!\!\!
\prod_{i=1}^{\nu}C \,M_s\,2^{h_i +1}R^{-h_i}
\leq C_{j,R}(1+ 2CM_s)^j
$$
where $C_{j,R}$ denotes a suitable positive constant depending on $j$ and $R$.
Since $\E(\int_t^T|M_s|^p\,ds)$ $<\infty$ for all $p$, one gets the desired integrability property for $G_2(s;\rho)$ as well.

So, we can actually write
$$
\partial^k_\rho
\E\big(\Gamma_2^{t,x,v}(\rho)\big)_{|_{\rho=0}}
= K e^{-r(T-t)}\sum_{j=0}^k\Big(\begin{array}{c}k\\j\end{array}\Big)\E\Big(\frac 1{\<M\>_{[t,T]}}\int_t^T \partial^{k-j}_\rho
G_1^{t,x,v}(s;\rho)\,\partial^{j}_\rho
G_2^{t,x,v}(s;\rho)\psi_s^{t,v}\,ds\Big)_{|_{\rho=0}}
$$
Now,
\begin{align*}
&\partial_\rho^{k-j}
G_1(s;\rho)_{|_{\rho=0}}
=(k-j)!\,G_{1,k-j}(s)
\quad\mbox{and}\\
&\partial_{\rho}^{j}G_2(s;\rho)_{|_{\rho=0}}=
j!\,\sum_{\nu=1}^{j}
\frac{N^{(2+\nu)}(D_0(s))}{\nu!}\!\!\!\!\!\!\!\!\!\!\!\sum_{\mbox{\scriptsize $ \begin{array}{c}h_i\geq 1\ \forall i\\ h_1 + ...+ h_{\nu}=j\end{array}$}}\!\!\!\!\!\!\!
\prod_{i=1}^{\nu}D_{h_i}(s)
\end{align*}
so that the statement actually holds.

Finally, for $n=0$ one has $u_0(t,x,v)=u(t,x,v;0)=\E(\Gamma^{t,x,v}_1)$ and for $n\geq 1$
$$
u_n(t,x,v)=\partial^n_\rho u_n(t,x,v;\rho)_{|_{\rho=0}}
=n\partial^{n-1}_\rho\E\big(\Gamma^{t,x,v}_2(\rho)\big)_{|_{\rho=0}}
$$
and the proof is completed.
\cvd

By using Proposition \ref{derivataGamma2tilde}, we can now immediately deduce a representation for the coefficients $u_0$, $u_1$, $u_2$. As already observed in  \cite{ScarlattiAntonelli}, from the numerical point of view these are the relevant objects which the attention has to be focalized to.

\begin{prop}\label{cor-u2}
Suppose that \textbf{(H)} and \textbf{(A)} hold and let $u_k$ be as in (\ref{uk}). Then the representation in terms of expectation for $u_0$ and $u_1$ both agree with the ones for $g_0$ and $g_1$, given by (\ref{gg0}) and (\ref{gg1}) respectively, whereas for $k=2$ one has
\begin{equation}
\label{u2finale}
u_2(t,x,v)=2Ke^{-r(T-t)}\mathbb{E}\Bigg(\frac1{\<M^{t,v}\>_T^{3/2}}
N^{(3)}(d_2)\bigg(t,x,\sqrt{\frac{\<M^{t,v}\>_T}{T-t}}\bigg)\,\ell^{t,v}_{[t,T]}\Bigg)
\end{equation}
where
$$
\ell_{[t,T]}^{t,v}
=\int_t^T
\psi_\alpha^{t,v}\int_t^\alpha f(v_\theta^{t,v})dB_\theta^1\,d\alpha
$$
the process $\psi^{t,v}$ being defined in (\ref{psi}).
\end{prop}

\emph{Proof.}
$u_0$ is trivially obtained putting $\rho=0$ in (\ref{alfa_beta}), and
this brings to the same representation of $g_0$ in (\ref{gg0}).
Regarding $u_1$, Proposition \ref{derivataGamma2tilde} gives
$$
\Xi^{t,x,v}_0(s)
=G_{1,0}^{t,v}(s)\,
G_{2,0}^{t,x,v}(s)
=[N^{(2)}(d_2)] \Big(t,x,\sqrt{\frac{\<M^{t,v}\>_{[t,T]}}{T-t}}\Big)
$$
and one gets $u_1=g_1$. Moreover,
$$
\Xi^{t,x,v}_1(s)
=G_{1,0}^{t,v}(s)\,
G_{2,1}^{t,x,v}(s)
=[N^{(3)}(d_2)] \Big(t,x,\sqrt{\frac{\<M^{t,v}\>_{[t,T]}}{T-t}}\Big)
\,\frac{\int_t^sf(v^{t,v}_u)\,dB^1_u}{\<M^{t,v}\>^{1/2}_{[t,T]}},
$$
hence formula (\ref{u2finale}) holds.
\cvd

We observe that the second order coefficient $u_2$ has a representation in terms of expectation which is really different from the one for $g_2$, as it follows by comparing (\ref{u2finale}) with (\ref{gg2}). Moreover, we would like to stress that representation (\ref{u2finale}) for $u_2$ is drastically simpler than expression (\ref{gg2}) for $g_2$: this permits us to perform numerical analysis with a much smaller effort.

\section{Series expansion M - from Malliavin calculus techniques}\label{sect-mall}

This section is devoted to the representation of the derivatives in terms of expectations by using the sensitivity representation  method due to the Malliavin calculus approach.

We first recall some notations and results from Malliavin calculus, and we refer e.g. to Nualart \cite{bib:n} and Bally \cite{bib:bally} for this topic.
Here, we perform the Malliavin calculus restricted to the time interval $[t,T]$, that is of interest for pricing options. So, once for all, we fix the starting instant $t\in [0,T)$ and the starting points $x,v\in\R$.

\smallskip

For $\ell\in\N$ and $p\geq 1$, we let $\D^{\ell,p}$ stand for the space of the
random variables which are $\ell$-times differentiable in the Malliavin sense in $L^{p}$. For $F\in\D^{\ell,p}$ and for a multi-index $\alpha$ such that $|\alpha|=j\leq \ell$ (that is, $\alpha=(\alpha _{1},\ldots ,\alpha _{j})\in
\{1,2\}^{j}$ -- recall that we are dealing with a Brownian motion on $\R^2$), $D^{\alpha }F$ denotes the Malliavin derivative of $F$
corresponding to the multi-index $\alpha.$ In particular, the notation
$DF=(D^1F,D^2F)$ stands for the first order Malliavin derivative of $F$ and as $i=1,2$, $D^i_sF$, $s\in[t,T]$, gives the derivative of $F$ in the direction of the $i$-th Brownian motion (roughly speaking, the derivative is done w.r.t. the infinitesimal increment $\Delta B^i_s$). As usual, we set $\D^{\ell,\infty}=\cap_{p\geq 1}\D^{\ell,p}$. Sometimes, for the sake of notation, we need to define $\D^{0,p}$ and $\D^{0,\infty}$. So, we put $\D^{0,p}=L^p(\Omega,\cl F_T,\P)$ and $\D^{0,\infty}=\cap_{p\geq 1}\D^{0,p}$.

Moreover, we let $Dom_2(\delta)$ denote the domain of the Skorohod integral, which is the adjoint operator of the Malliavin derivative in $\D^{1,2}$. And more generally, $Dom_p(\delta)$ denotes the set of the processes which are integrable in the Skorohod sense in $L^p$. As usual, we put $Dom_{\infty}(\delta)=\cap_{p}Dom_p(\delta)$.
For further use, we also consider the set $\L^{\ell,p}$ of the $2$-dimensional processes $g\in L^2([t,T]\times \Omega)$ such that $g_s\in\D^{\ell,p}$ for all $s\in[t,T]$ and for every multi-index $\alpha$ with $|\alpha|=j\leq \ell$ (that is, $\alpha\in\{1,2\}^j$) one has
$\|D^\alpha g\|_{L^2([t,T]^{j+1})}\in L^p(\Omega)$. Recall that $\L^{1,p}\subset   Dom_p(\delta)$. We define $\L^{\ell,\infty}=\cap_{p}\L^{\ell,p}$.

\smallskip

For $G\in\D^{\ell,p}$, $\psi\in Dom_p(\delta)$ and $g\in\L^{\ell,p}$, we recall the Malliavin-Sobolev norms
\begin{align*}
\|G\|_{\ell,p}^{p}
&=\E(|G|^{p})
+\sum_{j=1}^\ell\sum_{\alpha\,:\,|\alpha|=j}
\E\Big(\Big(\int_{[t,T]^j}|D^\alpha_{s_1\ldots s_j}G|^{2}\,ds_1\cdots ds_j\Big)^{p/2}\Big),\\
\|\psi\|_{\delta,p}^{p}
&=\E(|\delta(\psi)|^{p})
+\E\Big(\Big(\int_{[t,T]}|\psi_s|^{2}\,ds\Big)^{p/2}\Big),\\
\|g\|_{\L^{\ell,p}}^p
&=\E\Bigl(\Bigl(\int_t^T|g_s|^2\,ds\Bigr)^{p/2}\Bigr)
+\sum_{j=1}^\ell\sum_{\alpha\,:\,|\alpha|=j}
\E\Bigl(\Bigl(\int_{[t,T]^{j+1}}|D^\alpha_{s_1\ldots s_j}g_s|^2ds_1\cdots ds_j\,ds\Bigr)^{p/2}\Bigr).
\end{align*}
We also recall the following H\"older type inequalities:
for $\ell\in \N$ and $p\geq 1$, there exist constants $a_{\ell,p}>0$ and $b_{\ell,p}>0$ such that for every $\alpha,\beta> 1$ with $\frac 1\alpha+\frac 1\beta=1$  one has
\begin{align}
\|G_1G_2\|_{\ell,p}
&\leq a_{\ell,p}\,\|G_1\|_{\ell,p\alpha}\,\|G_2\|_{\ell,p\beta}\label{holder-D}\\
\|G\psi\|_{\delta,p}
&\leq b_{\ell,p}\,\|G\|_{\ell,p\alpha}\,\|\psi\|_{\delta,p\beta}\label{holder-delta}
\end{align}
for all $G_1,G_2,G\in\D^{\ell,\infty}$ and $\psi\in Dom_\infty(\delta)$.
Moreover, if $G\in\D^{\ell+1,\infty}$ and $g\in\L^{\ell,\infty}$ then $\int_t^T\<D_sG,g_s\>\,ds\in \D^{\ell,\infty}$ and for every $p$ there exists a constant $c_{\ell,p}$ such that for every $\alpha,\beta>1$ with $\frac 1\alpha+\frac 1\beta=1$ one has
\begin{equation}\label{holder-L}
\Big\|\int_t^T\<D_sG,g_s\>\,ds\Big\|_{\ell,p}\leq c_{\ell,p}\|G\|_{\ell+1,p\alpha}\|g\|_{\L^{\ell,p\beta}}.
\end{equation}
Let us remark that inequality (\ref{holder-L}) is perhaps less popular but it can be straightforwardly proved by using the same technique allowing one to get (\ref{holder-D}) and (\ref{holder-delta}).

\smallskip

To our purposes, we need an integration by parts formula which is the starting point of our results, and it is well known that Malliavin calculus produces integration by parts formulas. It is not the unique one but this gives formulas for the
series expansion coefficients which are feasible to be used in practice (see next Remark \ref{remIBP} for further details). And in order to proceed, we recall the notation $C^\infty_0(\R)$ to denote the set of the infinitely differentiable functions $\phi\,:\,\R\to\R$  with compact support. But we need also a condition,  slightly stronger than \textbf{(AS)}, which  is the following:

\begin{itemize}
\item[\textbf{(M)}] \emph{For every $t\in[0,T]$ and $v\in\R$ there exists a constant $C>0$ such that for any $q\ge1$,}
$$
\mathbb{E}\Big(\int_t^T|f(v_r^{t,v})|^{-q}dr\Big)\le C^q.
$$
\end{itemize}

Notice that, under \textbf{(H)} and \textbf{(M)}, one has $1/f(v_\cdot^{t,v})\in \L^{\ell,\infty}$ for all $\ell\in\N$ (actually, the right writing should be $g\in \L^{\ell,\infty}$ where $g_1\equiv 0$ and $g_2=1/f(v_\cdot^{t,v})$).
Then, we have:
\begin{lemma}\label{lemma1}
Assume that \textbf{(H)} and \textbf{(M)} hold and let $\ell\geq 1$.
Then the following integration by parts formula holds: for every $G\in\D^{\ell,\infty}$ one has
$$
\E\big(\phi'(\xi_T^{t,x,v})G\big)
=\E(\phi(\xi_T^{t,x,v})\,\H^{t,v}[G]\big)\quad \forall\ f\in C^\infty_0(\R)
$$
where the weight $\H^{t,v}[G]$ is given by
\begin{equation}\label{AS}
\H^{t,v}[G]
=\frac {(1-\rho^2)^{-1/2}}{T-t}\,\delta_2\Big(\frac G{f(v^{t,v}_\cdot)}\Big)
\end{equation}
in which $\delta_2$ stands for the Skorohod integral over $[t,T]$ w.r.t. the Brownian motion $B^2$.
Moreover, $\H^{t,v}[G]\in \D^{\ell-1,\infty}$.
\end{lemma}

\emph{Proof.} By the chain rule, for $s\in[t,T]$ one has
$$
D^2_s\phi(\xi_T^{t,x,v})=\phi'(\xi_T^{t,x,v})D^2_s\xi_T^{t,x,v}
=\phi'(\xi_T^{t,x,v})\sqrt{1-\rho^2}\,f(v^{t,v}_s)
$$
so that
$$
\phi'(\xi_T^{t,x,v})G
=\frac {(1-\rho^2)^{-1/2}}{T-t}\,\int_t^TD^2_s\phi(\xi_T^{t,x,v})\frac G{f(v^{t,v}_s)}\,ds.
$$
Since $G\in\D^{1,\infty}$ and $1/f(v^{t,v}_\cdot)\in Dom_\infty(\delta_2)$ then
$G/f(v^{t,v}_\cdot)\in Dom_\infty(\delta_2)$, so the duality relationship gives
$$
\E\big(\phi'(\xi_T^{t,x,v})G\big)
=\frac {(1-\rho^2)^{-1/2}}{T-t}\,\E\Big(\phi(\xi_T^{t,x,v})\delta_2\Big(\frac G{f(v^{t,v}_\cdot)}\Big)\Big).
$$
But one has also that $1/f(v^{t,v}_\cdot)\in \L^{\ell,\infty}$ and
$$
\delta_2\Big(\frac G{f(v^{t,v}_\cdot)}\Big)
=G\,\int_t^T \frac 1{f(v^{t,v}_\cdot)}\,dB^2_s-\int_t^T \frac{D^2_s G}{f(v^{t,v}_s)}\,ds,
$$
which says that $\H^{t,v}[G]\in\D^{\ell-1,\infty}$, and the statement is proved.
\cvd

Let us consider the following simple result.

\begin{lemma}\label{lemma2}
Let $\{T_n\}_n\subset \D^{\ell,\infty}$ be a sequence of r.v.'s 
such that for $p\geq 1$ and $R>0$,
$$
\sum_{n\geq 0}\|T_n\|_{\ell,p}\,\rho^n<\infty,\quad |\rho|<R.
$$
Then the r.v.
$$
T(\rho)=\sum_{n\geq 0}T_n\rho^n,\quad |\rho|<R,
$$
is well posed, belongs to $\D^{\ell,\infty}$ and for any multi-index $\alpha$ with $|\alpha|=j\leq \ell$,
$$
D^\alpha T(\rho)=\sum_{n\geq 0}D^\alpha T_n\, \rho^n,\quad|\rho|<R.
$$
Moreover, if $\psi\in Dom_\infty(\delta)$ then  $T(\rho)\psi\in Dom_\infty(\delta)$ and
$$
\delta(T(\rho)\psi)=\sum_{n=0}^\infty\delta(T_n\psi)\,\rho^n,\quad  |\rho|<R.
$$
\end{lemma}

\emph{Proof. }
The first assertion immediately follows by recalling that $\D^{\ell,p}$ is closed w.r.t. $\|\cdot\|_{\ell,p}$. As for the second one, since $Dom_p(\delta)$ is closed w.r.t. $\|\cdot\|_{\delta,p}$, it is enough to prove that $\sum_{n=0}^\infty \|T_n\psi\|_{\delta,p}\,\rho^n<\infty$ for $|\rho|<R$. By using (\ref{holder-delta}) one has
$\|T_n\psi\|_{\delta,p}\leq b_{1,p}\|T_n\|_{1,\bar p}\|\psi\|_{\delta,\bar q}$ for suitable $\bar p,\bar q>p$, so that
$$
\sum_{n=0}^\infty \|T_n\psi\|_{\delta,p}\,\rho^n
\leq c_p\|\psi\|_{\delta,\bar q}\,
\sum_{n=0}^\infty \|T_n\|_{\delta,\bar p}\,\rho^n<\infty,\quad |\rho|<R.
$$
\cvd

We can now state the key result of this section.

\begin{lemma}\label{lemma-mall}
Assume that \textbf{(H)} and \textbf{(M)} hold. For $k\geq 0$, set
\begin{equation}\label{lambda-k}
\mbox{$\Lambda_0^{t,x,v}(\rho)=1$ and for $k\geq 1$, $\Lambda^{t,v}_k(\rho)=\H^{t,v}[G^{t,v}_\rho\Lambda^{t,v}_{k-1}(\rho)]
+\partial_\rho\Lambda_{k-1}^{t,v}(\rho)$.}
\end{equation}
where $\H^{t,v}[\cdot]$ is given in (\ref{AS}) and
\begin{equation}\label{Gro}
G^{t,v}_\rho=
\int_t^Tf(v^{t,v}_u)\,dB_u^1-\frac{\rho}{\sqrt{1-\rho^2}}\int_t^Tf(v^{t,v}_u)\,dB_u^2
\end{equation}
Then for every $k\in\N$,
$\Lambda_k^{t,v}(\rho)$ is well posed, belongs to $\D^{\ell,\infty}$ for all $\ell\in\N$ and can be represented as
$$
\Lambda_k^{t,v}(\rho)=\sum_{n=0}^\infty \Lambda^{t,v}_{k,n}\,\rho^n,
$$
with $\{\Lambda^{t,v}_{k,n}\}_{n}$ given by
\begin{equation}\label{lambda-kn-bis}
\begin{array}{rl}
\bullet\quad
k=0,\,n\geq 0:&
\mbox{$\Lambda_{0,0}^{t,v}=1$ and  $\Lambda^{t,v}_{0,n}=0$ for all $n\geq 1$},\smallskip\\
\bullet\quad
k\geq 1,\,n\geq 0:
& \Lambda_{k,n}^{t,v}
=\displaystyle
\frac 1{T-t}\Big[\sum_{\ell\,:\,2\ell\leq n}
\frac{(2\ell)!}{(\ell!)^2\,4^\ell}
\delta_2\Big(\frac{U^{t,v}\Lambda^{t,v}_{k-1,n-2\ell}}{f(v^{t,v}_\cdot)}\Big)+\smallskip\\
&\qquad\qquad\qquad-\!\!\!\!
\displaystyle
\sum_{\ell\,:\,2\ell+1\leq n}
\delta_2\Big(\frac{V^{t,v}\Lambda^{t,v}_{k-1,n-2\ell-1}}{f(v^{t,v}_\cdot)}\Big)\Big]
+(n+1)\Lambda_{k-1,n+1}^{t,v}
\end{array}
\end{equation}
in which
$$
U^{t,v}=\int_t^Tf(v_s^{t,v})dB^1_s \quad\mbox{and}\quad V^{t,v}=\int_t^Tf(v_s^{t,v})dB^2_s.
$$

\end{lemma}

\emph{Proof.}
We prove that for every $k\geq 0$, there exists a sequence $\{\Lambda_{k,n}^{t,v}\}_n$ independent of $\rho$  such that
\begin{equation}\label{ind}
\begin{array}{l}
\displaystyle
\bullet\quad
\{\Lambda_{k,n}^{t,v}\}_n\subset\D^{\ell,\infty} \mbox{ for every } \ell\in\N;\smallskip\\
\displaystyle
\bullet\quad
\sum_{n=0}^\infty \|\Lambda_{k,n}^{t,v}\|_{\ell,p}\,\rho^n<\infty,\quad  \mbox{$|\rho|<1$,
$p\geq 1$ and $\ell\in\N$;}
\smallskip\\
\displaystyle
\bullet\quad
\Lambda_k^{t,v}(\rho)=\sum_{n=0}^\infty \Lambda_{k,n}^{t,v}\,\rho^n,\quad |\rho|<1.
\end{array}
\end{equation}
If (\ref{ind}) holds, we can apply Lemma \ref{lemma2}, so that for $|\rho|<1$ we get  $\Lambda_k^{t,v}(\rho)\in\D^{\ell,\infty}$ for all $\ell$. And as a consequence of the power series expansion, $\Lambda_k^{t,v}(\rho)$ is continuously differentiable w.r.t. $\rho$.

We prove now (\ref{ind})  by induction.

If $k=0$, (\ref{ind}) trivially holds. So, we assume that (\ref{ind})  is true for $k-1$, $k\geq 1$, and we prove that (\ref{ind}) holds for $k$ as well.
For the sake of simplicity,
set $U^{t,v}=\int_t^Tf(v_s^{t,v})dB^1_s$ and $V^{t,v}=\int_t^Tf(v_s^{t,v})dB^2_s$, so that $G^{t,v}_\rho=U^{t,v}-\rho(1-\rho^2)^{-1/2}V^{t,v}$. Notice that under \textbf{(H)} one has $U^{t,v},V^{t,v}\in\D^{\ell,\infty}$ for all $\ell$. From (\ref{AS}) we have
\begin{align*}
\H^{t,v}[G^{t,v}_\rho\Lambda_{k-1}^{t,v}(\rho)]
&=\frac{(1-\rho^2)^{-1/2}}{T-t}\delta_2\Big(\frac{G^{t,v}_\rho\Lambda_{k-1}^{t,v}(\rho)}{f(v^{t,v}_\cdot)}\Big)\\
&=\frac{1}{T-t}\Big[(1-\rho^2)^{-1/2}
\delta_2\Big(\frac{U^{t,v}\Lambda_{k-1}^{t,v}(\rho)}{f(v^{t,v}_\cdot)}\Big)
-\rho(1-\rho^2)^{-1}\delta_2\Big(\frac{V^{t,v}\Lambda_{k-1}^{t,v}(\rho)}{f(v^{t,v}_\cdot)}\Big)
\Big]
\end{align*}
We prove now that one actually has $U^{t,v}\Lambda_{k-1}^{t,v}(\rho)/f(v^{t,v}_\cdot)$, $V^{t,v}\Lambda_{k-1}^{t,v}(\rho)/f(v^{t,v}_\cdot)\in Dom_\infty(\delta_2)$. So, we set
$\psi$ such that $\psi_1(s)=0$ and $\psi_2(s)=1/f(v^{t,v}_s)$. We stress that \textbf{(H)} and \textbf{(M)} both give $\psi\in Dom_{\infty}(\delta)$. Since (\ref{ind}) holds for $k-1$,
we can write
$$
\frac{U^{t,v}\Lambda_{k-1}^{t,v}(\rho)}{f(v^{t,v}_s)}
=\sum_{n=0}^\infty U^{t,v}\Lambda_{k-1,n}^{t,v}\psi_2(s)\,\rho^n,\quad |\rho|<1
$$
where, for every $\ell\in\N$, $\{\Lambda_{k-1,n}^{t,v}\}_n\subset\D^{\ell,\infty}$ and for any $p$ one has
$\sum_{n=0}^\infty \|\Lambda_{k-1,n}^{t,v}\|_{\ell,p}\,\rho^n<\infty$ if $|\rho|<1$.
So, by Lemma \ref{lemma2}, it is sufficient to prove that $\sum_{n=0}^\infty \|U^{t,v}\Lambda_{k-1,n}^{t,v}\|_{1,p}\,\rho^n<\infty$ for $|\rho|<1$. From (\ref{holder-D}), there exists $\bar p,\bar q>p$ such that $\|U^{t,v}\Lambda_{k-1,n}^{t,v}\|_{1,p}\leq a_{1,p}\|U^{t,v}\|_{1,\bar p}\,\|\Lambda_{k-1,n}^{t,v}\|_{1,\bar q}$, so that
$$
\sum_{n=0}^\infty \|U^{t,v}\Lambda_{k-1,n}^{t,v}\|_{1,p}\,\rho^n
\leq a_{1,p}\|U^{t,v}\|_{1,\bar p}\,\sum_{n=0}^\infty \|\Lambda_{k-1,n}^{t,v}\|_{1,\bar q}\,\rho^n<\infty.
$$
This gives $U^{t,v}\Lambda_{k-1}^{t,v}(\rho)/f(v^{t,v}_\cdot)\in Dom_\infty(\delta_2)$ and moreover,
$$
\delta_2\Big(\frac{U^{t,v}\Lambda_{k-1}^{t,v}(\rho)}{f(v^{t,v}_\cdot)}\Big)
=\sum_{n=0}^\infty \delta_2\Big(\frac{U^{t,v}\Lambda_{k-1,n}^{t,v}}{f(v^{t,v}_\cdot)}\Big)\,\rho^n.
$$
Similarly, we have that $V^{t,v}\Lambda_{k-1}^{t,v}(\rho)/f(v^{t,v}_\cdot)\in Dom_\infty(\delta_2)$ and
$$
\delta_2\Big(\frac{V^{t,v}\Lambda_{k-1}^{t,v}(\rho)}{f(v^{t,v}_\cdot)}\Big)
=\sum_{n=0}^\infty \delta_2\Big(\frac{V^{t,v}\Lambda_{k-1,n}^{t,v}}{f(v^{t,v}_\cdot)}\Big)\,\rho^n.
$$
So, we can write
$$
\H^{t,v}[G^{t,v}_\rho\Lambda_{k-1}^{t,v}(\rho)]
=\frac{1}{T-t}\sum_{n=0}^\infty\Big[
(1-\rho^2)^{-1/2}
\delta_2\Big(\frac{U^{t,v}\Lambda_{k-1,n}^{t,v}}{f(v^{t,v}_\cdot)}\Big)
-\rho(1-\rho^2)^{-1}\delta_2\Big(\frac{V^{t,v}\Lambda_{k-1,n}^{t,v}}{f(v^{t,v}_\cdot)}\Big)\Big]\rho^n.
$$
Now, since $(1-\rho^2)^{-1/2}=\sum_{n\geq 0}\frac{(2n)!}{(n!)^2\,4^n}\,\rho^{2n}$ and $\rho(1-\rho)^{-1} =\sum_{n\geq 0}\rho^{2n+1},$
straightforward computations give
\begin{align*}
\H^{t,v}[G^{t,v}_\rho\Lambda_{k-1}^{t,v}(\rho)]
&=\frac 1{T-t}\sum_{n= 0}^\infty\Big[\sum_{\ell\geq 0\,:\,2\ell\leq n}
\frac{(2\ell)!}{(\ell!)^2\,4^\ell}
\delta_2\Big(\frac{U^{t,v}\Lambda_{k-1,n-2\ell}}{f(v_\cdot)}\Big)+\\
&\qquad\qquad \qquad\quad
-\!\!\!\!\!\sum_{\ell\geq 0\,:\,2\ell+1\leq n}\delta_2\Big(\frac{V^{t,v}\Lambda_{k-1,n-2\ell-1}}{f(v_\cdot)}\Big)\Big]\,\rho^n.
\end{align*}
Moreover, one has
$$
\partial_\rho\Lambda_{k-1}^{t,v}(\rho)=\sum_{n= 0}^\infty (n+1)\Lambda_{k-1,n+1}^{t,v}\rho^n,\quad |\rho|<1,
$$
with
$$
\sum_{n= 0}^\infty (n+1)\|\Lambda_{k-1,n+1}^{t,v}\|_{\ell,p}\rho^n<\infty,\quad |\rho|<1
$$
by the square-test. Therefore, by resuming we get
$\Lambda_k^{t,v}(\rho)=\sum_{n=0}^\infty \Lambda_{k,n}^{t,v}\,\rho^n$
with
$$
\begin{array}{rl}
\Lambda_{k,n}^{t,v}
=&
\displaystyle
\frac 1{T-t}\Big[\sum_{\ell\geq 0\,:\,2\ell\leq n}
\frac{(2\ell)!}{(\ell!)^2\,4^\ell}
\delta_2\Big(\frac{U^{t,v}\Lambda_{k-1,n-2\ell}}{f(v^{t,v}_\cdot)}\Big)+\smallskip\\
&\qquad\qquad-\displaystyle
\!\!\!\!\sum_{\ell\geq 0\,:\,2\ell+1\leq n}\delta_2\Big(\frac{V^{t,v}\Lambda_{k-1,n-2\ell-1}}{f(v^{t,v}_\cdot)}\Big)\Big]+(n+1)\Lambda_{k-1,n+1}^{t,v},
\end{array}
$$
which gives formula (\ref{lambda-kn-bis}). Now, in order to conclude (and to justify the above change of order of the series indexes), it remains to show that
$$
\sum_{n= 0}^\infty \|\Lambda_{k,n}^{t,v}\|_{\ell,p}\,\rho^n<\infty,\quad |\rho|<1,
$$
for all $\ell$ and $p$. Notice that, by re-adjusting the indexes of the series, it is sufficient to prove that if $|\rho|<1$,
$$
\sum_{n= 0}^\infty
\Big\|\delta_2\Big(\frac{U^{t,v}\Lambda_{k-1,n}^{t,v}}{f(v^{t,v}_\cdot)}\Big)\Big\|_{\ell,p}\,\rho^n<\infty
\quad\mbox{and}\quad
\sum_{n= 0}^\infty
\Big\|\delta_2\Big(\frac{V^{t,v}\Lambda_{k-1,n}^{t,v}}{f(v^{t,v}_\cdot)}\Big)\Big\|_{\ell,p}\,\rho^n<\infty $$
(recall that $\sum_{n= 0}^\infty (n+1)\|\Lambda_{k-1,n+1}^{t,v}\|_{\ell,p}\rho^n<\infty$,  $|\rho|<1$).
Consider the first series and, for the sake of simplicity, let us set $g$ as a $2$-dimensional process whose first component is null and its second one is given by $U^{t,v}/f(v^{t,v}_\cdot)$. Thanks to \textbf{(H)} and \textbf{(M)} one gets $g\in \L^{\ell,\infty}$ for all $\ell$. Moreover,
$$
\delta_2\Big(\frac{U^{t,v}\Lambda_{k-1,n}^{t,v}}{f(v^{t,v}_\cdot)}\Big)
=\Lambda_{k-1,n}\delta(g)-\int_t^T \<D_s \Lambda_{k-1,n}^{t,v},g_s\>ds,
$$
so that
\begin{align*}
\Big\|\delta_2\Big(\frac{U^{t,v}\Lambda^{t,v}_{k-1,n}}{f(v^{t,v}_\cdot)}\Big)\Big\|_{\ell,p}
&\leq
\|\Lambda^{t,v}_{k-1,n}\delta(g)\|_{\ell,p}+\Big\|\int_t^T \<D_s \Lambda^{t,v}_{k-1,n},g_s\>ds\Big\|_{\ell,p}\\
&\leq C_{\ell,p}\Big(\|\Lambda^{t,v}_{k-1,n}\|_{\ell,\bar p}\,\|\delta(g)\|_{\ell,\bar q}+
\|\Lambda^{t,v}_{k-1,n}\|_{\ell+1,\bar p}\,\|g\|_{\L^{\ell,p}}\Big)\\
&\leq C_{\ell,p}\Big(\|\delta(g)\|_{\ell,\bar q}+
\|g\|_{\L^{\ell,\bar q}}\Big)\|\Lambda^{t,v}_{k-1,n}\|_{\ell+1,\bar p},
\end{align*}
for suitable $\bar p,\bar q>p$ and a positive constant $C_{\ell,p}$ coming from the use of inequalities (\ref{holder-D}) and (\ref{holder-L}). By recalling that, by induction, one has $\sum_{n= 0}^\infty \|\Lambda^{t,v}_{k-1,n}\|_{\ell+1,\bar p}\,\rho^n<\infty$, one gets
$\sum_{n= 0}^\infty
\|\delta_2(\frac{U^{t,v}\Lambda^{t,v}_{k-1,n}}{f(v^{t,v}_\cdot)})\|_{\ell,p}\,\rho^n<\infty$. Similarly, one proves that
$\sum_{n= 0}^\infty
\|\delta_2(\frac{V^{t,v}\Lambda^{t,v}_{k-1,n}}{f(v^{t,v}_\cdot)})\|_{\ell,p}\,\rho^n<\infty$, and the statement is now completely proved.
\cvd

We are now ready to prove our main result.

\begin{thm}\label{thm-mall}
Assume that \textbf{(H)} and \textbf{(M)} hold. Let $\phi$ denote a Borel function with exponential growth, that is $|\phi(x)|\leq C(1+e^{\lambda x})$, for some $C,\lambda>0$.
Then for every $k\geq 1$ one has
\begin{equation}\label{repr-mall}
\partial^k_\rho\E\big(\phi(\xi_T^{t,x,v})\Big)
=\E\big(\phi(\xi_T^{t,x,v})\,\Lambda_k^{t,v}(\rho)\big)
\end{equation}
where $\Lambda_k^{t,v}(\rho)$ is given by (\ref{lambda-k}).
In particular, one has
$$
\partial_\rho^k \E\big(\phi(\xi_T^{t,x,v})\big)_{|_{\rho=0}}
=\E\big(\phi(\hat\xi_T^{t,x,v})\Lambda_{k,0}^{t,v}\big)
$$
where $\Lambda^{t,v}_{k,0}$ is the starting point of the sequence $\{\Lambda^{t,v}_{k,n}\}_{n}$ defined through (\ref{lambda-kn-bis}).

\end{thm}

\emph{Proof.}
We first assume that $\phi\in C^\infty_0$, the general case to be treated by regularizing arguments.

For $k=0$, (\ref{repr-mall}) trivially holds. By induction, we assume that (\ref{repr-mall}) holds for $k-1$, $k\geq 1$, and we prove that (\ref{repr-mall}) holds for $k$ as well.
In fact, by recalling that $\partial_\rho\xi_T^{t,x,v}=G_\rho^{t,v}$, one has
\begin{align*}
\partial^k_\rho\E\big(\phi(\xi_T^{t,x,v})\Big)
&=\partial_\rho \E\big(\phi(\xi_T^{t,x,v})\,\Lambda_{k-1}^{t,v}(\rho)\big)\\
&=\E\big(\phi'(\xi_T^{t,x,v})\partial_\rho \xi_T^{t,x,v} \,\Lambda_{k-1}^{t,v}(\rho)\big)
+\E\big(\phi(\xi_T^{t,x,v})\,\partial_\rho\Lambda_{k-1}^{t,v}(\rho)\big)\\
&=\E\Big(\phi(\xi_T^{t,x,v})\Big(\H^{t,v}[\partial_\rho \xi_T^{t,x,v} \,\Lambda_{k-1}^{t,v}(\rho)]
+\partial_\rho\Lambda_{k-1}^{t,v}(\rho)\Big)\Big)\\
&\equiv\E\Big(\phi(\xi_T^{t,x,v})\Lambda_k^{t,v}(\rho)\Big)
\end{align*}
and the above interchange between the operators $\E$ and $\partial_\rho$ is due to the fact that from Lemma \ref{lemma-mall} one has $\Lambda^{t,v}_k(\rho)\in L^p$ for all $k$ and $p$.

Now, in order to deal with a function $\phi$ which is simply Borel measurable and with exponential growth, we use arguments similar to the ones developed in Fourni\'e \emph{et al.} \cite{Fournie} (see the proof of Proposition 3.2, at page 400) by approximating $\phi$ with a sequence $\{\phi_n\}_n\subset C_0^\infty$. We first notice that $\phi(\xi^{t,x,v}_s)\equiv h(S_s^{t,x,v})$ with $h$ having polynomial growth, so that  $\sup_{|\rho|<R}\sup_{t\leq s\leq T}|\phi(\xi^{t,x,x}_s)|\in L^p$ for every $p$. So, it is enough  to prove that
$\sup_{|\rho|\leq R}|\Lambda_k^{t,v}(\rho)|\in L^p$ for every $R<1$. And in fact, since
$$
\sup_{|\rho|\leq R}|\Lambda_k^{t,v}(\rho)|
\leq \sum_{n=0}^\infty |\Lambda_{k,n}^{t,v}|\,R^n
$$
one gets
$$
\big\|\sup_{|\rho|\leq R}|\Lambda_k^{t,v}(\rho)|\big\|_p
\leq \sum_{n=0}^\infty \|\Lambda_{k,n}^{t,v}\|_{p}\,R^n
\leq \sum_{n=0}^\infty \|\Lambda_{k,n}^{t,v}\|_{1,p}\,R^n
$$
which is finite, as already seen in the proof of Lemma \ref{lemma-mall}, see (\ref{ind}).
Finally, since $\Lambda^{t,v}_k(0)=\Lambda^{t,v}_{k,0}$, last statement holds as well.
\cvd

As a particular case, useful for practical purposes, we can state the following

\begin{prop}\label{prop-mall}
Assume that \textbf{(H)} and \textbf{(M)} hold.
Let $\phi$ denote a Borel measurable function with exponential growth. Then one has
$$
\partial_\rho\E\big(\phi(\xi_T^{t,x,v})\big)_{|_{\rho=0}}
=\E\big(\phi(\hat\xi_T^{t,x,v})\Lambda_{1,0}^{t,v}\big)
\quad\mbox{and}\quad
\partial^2_\rho\E\big(\phi(\xi_T^{t,x,v})\big)_{|_{\rho=0}}
=\E\big(\phi(\hat\xi_T^{t,x,v})\Lambda_{2,0}^{t,v}\big)
$$
where
\begin{align*}
\Lambda^{t,v}_{1,0}
=&\frac 1{T-t}\,U^{t,v}Z^{t,v},\\
\Lambda^{t,v}_{2,0}
=&\frac {(U^{t,v})^2}{(T-t)^2}\Big((Z^{t,v})^2-\int_t^T |f(v_s^{t,v})|^{-2}
\,ds\Big)
-\frac 1{T-t}\,V^{t,v}Z^{t,v}+1,
\end{align*}
in which
$$
U^{t,v}=\int_t^Tf(v_s^{t,v})dB^1_s,\quad
V^{t,v}=\int_t^Tf(v_s^{t,v})dB^2_s,\quad
Z^{t,v}=\int_t^T\frac 1{f(v_s^{t,v})}dB^2_s.
$$
\end{prop}

\emph{Proof.}
We apply Theorem \ref{thm-mall}: by using (\ref{lambda-kn-bis}), we only have to compute $\Lambda_{1,0}^{t,v}$ and $\Lambda_{2,0}^{t,v}$, for which we need also $\Lambda_{1,1}^{t,v}$. Since $U^{t,v}$ is independent of $B^2$ and
$\Lambda_{0,0}^{t,v}=1$, we have
$$
\Lambda_{1,0}^{t,v}
=\frac 1{T-t}U^{t,v}\int_t^T\frac 1{f(v^{t,v}_s)}\,dB^2_s
=\frac 1{T-t}\,U^{t,v}Z^{t,v}
$$
As for $\Lambda_{1,1}^{t,v}$, by recalling that $\Lambda_{0,1}^{t,v}=0$, (\ref{lambda-kn-bis}) gives
\begin{align*}
\Lambda_{1,1}^{t,v}
&=-\frac 1{T-t}\delta_2\Big(\frac{V^{t,v}}{f(v^{t,v}_\cdot)}\Big)
=-\frac 1{T-t}\,\Big(V^{t,v}\int_t^T\frac 1{f(v^{t,v}_s)}\,dB^2_s-\int_t^T\frac{D^2_s V^{t,v}} {f(v^{t,v}_s)}\,ds\Big)\\
&=-\frac 1{T-t}\,V^{t,v}Z^{t,v}+1
\end{align*}
because $D^2_sV^{t,v}=f(v_s^{t,v})$. Finally, since $D^2_sZ^{t,v}=1/f(v^{t,v}_s)$ one has
\begin{align*}
\Lambda_{2,0}^{t,v}
=&\frac 1{(T-t)^2}\delta_2\Big(\frac{(U^{t,v})^2Z^{t,v}}{f(v^{t,v}_\cdot)}\Big)
-\frac 1{T-t}\,V^{t,v}Z^{t,v}+1\\
=&\frac {(U^{t,v})^2}{(T-t)^2}\Big(Z^{t,v}\int_t^T\frac{1}{f(v^{t,v}_s)}\,dB^2_s
-\int_t^T\frac{D^2_sZ^{t,v}}{f(v^{t,v}_s)}\,ds\Big)
-\frac 1{T-t}\,V^{t,v}Z^{t,v}+1\\
=&\frac {(U^{t,v})^2}{(T-t)^2}\Big((Z^{t,v})^2-\int_t^T\frac 1{f(v_s^{t,v})^2}\,ds\Big)
-\frac 1{T-t}\,V^{t,v}Z^{t,v}+1.
\end{align*}
\cvd

\begin{remark}\label{remIBP}
The formulas in Proposition \ref{prop-mall} follow from the Malliavin integration by parts formula in Lemma \ref{lemma1}, which in turn is stated by means of the noise given by $B^2$. This gives rise to simple formulas, mainly because $B^2$ does not appear in the dynamics for the volatility process. It is worth to be said that other integration by parts formulas can be provided, also by considering both noises $B^1$ and $B^2$, and  results similar to Proposition \ref{prop-mall} can be stated. For example, we have found several weights that involve all the noises and lead again to an expression for the derivatives of order one and two, evaluated in $\rho=0$, written in terms of some suitable weights. But such resulting weights are really complicated to write down and then unfeasible to be used in practice. However, as we will see in next Section \ref{sect-numerics}, for numerical purposes the weights in Proposition \ref{prop-mall} work very efficiently, and this is a further reason to drop complications.
\end{remark}

\begin{remark}\label{jumps}
The formulas for the weights continue to hold if jumps are suitably inserted in the model. For example, Al\`os, L\'eon and Vives \cite{Alos3} consider a jump-diffusion model for the log-returns by inserting in the dynamics (\ref{din-xi}) an independent compount Poisson process (actually, in that paper the model for the volatility is more general -- it is not a diffusion process, neither Markovian, but it is measurable w.r.t. the noise $B^1$) and they prove that a Hull and White type representation for the call price again holds (see Theorem 4.1 therein). In our framework, one would have
\begin{equation}\label{din-xi-jumps}
\xi_s= x+\int_t^s\Big(r-m\lambda-\frac12f^2(v_u)\Big)du+\int_t^sf(v_u)\big(\rho\,dB_u^1+\sqrt{1-\rho^2}\, dB_u^2\big)+\sum_{n=1}^{N_s}\Delta_n,
\end{equation}
where $v$ satisfies (\ref{din2}), $N$ stands for a Poisson process with intensity $\lambda>0$, the sequence of the jumps $\{\Delta_n\}_n$ is modeled by i.i.d. r.v's and the random sources $B$, $N$ and $\{\Delta_n\}_n$ are independent. Since we consider the evolution of the log-returns under the risk neutral measure, in (\ref{din-xi-jumps}) one has $m=\int(e^z-1)\mu(dz)$, $\mu$ denoting the common law of the $\Delta_n$'s. Now, if we assume that $e^{\Delta_1}\in L^p$ for every $p$ (and this holds in cases of interest, e.g. when $e^{\Delta_1}$ follows a log-normal or an exponential law), then it is easy to see that Proposition \ref{prop-mall} continues to hold with $\xi$ fulfilling (\ref{din-xi-jumps}), as a consequence of the fact that the noise from the jumps is independent of the Brownian noise. This is done e.g. in Bally, Bavouzet-Morel and Messaud \cite{bib:bbm}, where a Malliavin calculus in the direction of the jumps is also developed, possibly giving other integration by parts formulas.
\end{remark}

\begin{remark}\label{loc}
In practice, it is well known that the Malliavin representation for sensitivities leads to a problem of high variance. A standard variance reduction technique uses localizing functions, as follows.
Let $h$ denote a payoff-function, assumed to be Borel measurable and with polynomial growth. By Proposition \ref{prop-mall}, for a suitable regular and with bounded derivatives function $\Phi$, we can write
$$
\partial_\rho\mathbb{E}\big(h(S_T^{ t,x,v})\big)
=\mathbb{E}\big((h-\Phi)(S_T^{t,x,v})\Lambda_{1}^{t,v}(\rho)\big)
+\mathbb{E}\big(\partial_{x}\Phi(S_T^{t,x,v})\partial_{\rho}S_T^{t,x,v}\big)
$$
and similarly,
$$
\partial^2_\rho\mathbb{E}(h(S_T^{ t,x,v}))
=\mathbb{E}\big((h-\Phi)(S_T^{t,x,v})\Lambda_2^{t,v}(\rho)\big)
+\mathbb{E}\big(
\partial^2_{x}\Phi(S_T^{t,x,v})(\partial_{\rho}S_T^{t,x,v})^2+
\partial_{x}\Phi(S_T^{t,x,v})\partial^2_{\rho}S_T^{t,x,v}\big).
$$
So, the idea is to take $\Phi$ such that $h-\Phi$ cannot assume values too different each other. This procedure permits to decrease the contribution given by high fluctuations of the weights $\Lambda_1^{t,v}(\rho)$ and $\Lambda_2^{t,v}(\rho)$ and, as a consequence, it improves the numerical results and reduces the variance.

The literature on these arguments already contain many different possibilities for the localizing function $\Phi$. We need here $\Phi\in C^2$ that allows to work efficiently for the second order representation. So, in our numerical analysis we consider a slightly modification (more regularity) of the localizing function for call options developed in Bavouzet-Morel and Messaud \cite{BavouzetMessaoud}. In particular we take
\begin{equation}\label{Loc_Pol}
\Phi(y)=\int_{-\infty}^{y}\int_{-\infty}^wB_{\delta}(z)dz\,dw \mbox{ with }
B_{\delta}(z)=
\left\{
\begin{array}{ll}
0 & \mbox {if } z \le K-\delta\\
\displaystyle-3\,\frac {(z-K)^2}{4\delta^3}+\frac {3} {4\delta}&\mbox{if } z\in[K-\delta, K+\delta]\\
0 & \mbox{ if } z\ge K+\delta
\end{array}
\right.
\end{equation}
\end{remark}

\section{Numerical examples}\label{sect-numerics}

We study here the numerical behavior of the series coefficient representations:
we use the results of sections \ref{sect-alos} and \ref{sect-mall} to numerically compute by means of a Monte Carlo method the call price through its series expansions up to order 1 and 2 and
we compare the results with the ones in Antonelli and Scarlatti \cite{ScarlattiAntonelli} (where it is clear that only the terms up to order 2 are numerically relevant).

The tests in \cite{ScarlattiAntonelli} are related to the most popular stochastic volatility models:
\begin{itemize}
\item
the Hull and White model \cite{HullandWhite}, in which
$$
\mu(v)=\mu\,v,\quad \eta(v)=c\,v,\quad f(v)=v;
$$
\item
the Stein and Stein model \cite{bib:stein}, in which
$$
\mu(v)=b(a-v),\quad \eta(v)=c,\quad f(v)=v;
$$
\item
the Heston  model \cite{heston}, in which
$$
\mu(v)=b(a-v),\quad \eta(v)=c\,\sqrt v,\quad f(v)=\sqrt v.
$$
\end{itemize}

These are models for which the validity of the requirements \textbf{(H)}, \textbf{(A)} and \textbf{(M)} deserves some comments (we recall that \textbf{(AS)} always holds, as proved in \cite{ScarlattiAntonelli}, Appendix A). First of all, the fact that \textbf{(H)} is not fulfilled for the Heston model is not really a problem (as already observed in
\cite{ScarlattiAntonelli}, Section 5.1). Concerning  \textbf{(A)} and \textbf{(M)}, they  are not generally verified by the models we are concerned in.
For example, \textbf{(A)} always fails. This is not really a problem in the Hull and White model, since a transformation can be done in order to get $f(v)=const>0$, but it does become an unpleasant fact for the other models. However, in order that \textbf{(A)} and therefore \textbf{(M)} hold, one could think to perturb a little bit the model. Since problems come from a function of the type $g(v)=v^{\alpha}$, $\alpha=1,1/2$,
one could replace $g$ with $g_\varepsilon(v)=\sqrt{v^{2\alpha} + \varepsilon}$.
%
Such a perturbation procedure has been exploited in our numerical simulations concerning the Stein and Stein model and the Heston model. In any case, we have set $\varepsilon=10^{-5}$.

\smallskip

Let us now enter in the details of the simulation procedures.

\smallskip

We use an Euler approximation scheme to simulate the dynamics of the diffusions over $[0,T]$ (in all tests, we have set $t=0$). For the simulation of the volatility process in the Heston model (following a CIR process), we have done a further approximation: in the approximating scheme, the diffusion coefficient $\eta(v)=\sqrt{v}$ of the volatility process is replaced by
$\eta_{\gamma}(v)=\sqrt{|v| + \gamma}$. In practice, we have set $\gamma=10^{-5}$.
It is worth to be said that the use of more accurate schemes for the simulation of the CIR process could be used, thus avoiding this further complication. For example, Alfonsi \cite{bib:alfonsi} has developed a second order approximation scheme for the Heston model, but it deserves to be modified in order to simulate the functionals we need in our formulas. However, as it follows from the numerical tests, our simulation scheme behaves efficiently.

\smallskip

So, in our experiments we use the above Euler approximation scheme to simulate the dynamics of the diffusion processes; moreover, a Monte Carlo algorithm is developed to compute the expectations giving the coefficients of the power series. We divided the time interval $[0,T]$ in $N=500$ parts and we have performed $10^4$ simulations, and this means that we have used a standard number of parameters for the Euler scheme and for the Monte Carlo replications.

The numerical results refer to the percentage errors from the different expansions up to order 1 and 2. For the sake of clearness, we precise that the ``percentage error'' is given by
\begin{equation*}
\frac{|\hat p- p|}{ p}\times 100.
\end{equation*}
where $p$ denotes the true option price (or a benchmark value for it) and $\hat p$ stands for the approximation for $p$.

The tables give the percentage errors in our numerical experiments from the various methods and are labeled as follows:
\begin{itemize}
\item \virg AS' displays the results in Antonelli and Scarlatti \cite{ScarlattiAntonelli};
\item \virg ExpA-1' and \virg ExpA-2' refer to  the Taylor series expansion up to the first and the second order  respectively, obtained from the results developed in Section \ref{sect-alos};
\item \virg ExpM-1' and \virg ExpM-2' relate to the Taylor series expansion up to the first and to the second order respectively,  obtained in Section \ref{sect-mall} and refined with the localizing function (\ref{Loc_Pol}), in which $\delta=10^{-2}K$, $K$ denoting the (varying) strike price.
\end{itemize}

We stress that the results labeled `AS' come from the original paper \cite{ScarlattiAntonelli} (see Section 5 therein) and we recall that they are computed through the first order  expansion by means of a further approximation for the coefficients $g_0$ and $g_1$ (see (\ref{gg0}) and (\ref{gg1}) respectively): $g_0$ and $g_1$ are changed with $\overline{g}_0$ and $\overline{g}_1$ obtained by replacing $\<M^{t,v}\>_{[t,T]}$  with $\mathbb{E}(\<M^{t,v}\>_{[t,T]})$. It is clear that $\overline{g}_0$ and $\overline{g}_1$ give a ready first-order closed-formula approximation for the call price, which is very accurate.
So, the method `ExpA-1' gives the first order expansion from the Antonelli and Scarlatti approach deprived of any further approximation, and we will see that is works better in several cases.

\smallskip

The tables are collected in Section  \ref{sect-tables}.
They refer to varying values of $\rho$ and of the strike price $K$. As for the other parameters, the choice is done according to the model and can be detailed as follows.

\begin{itemize}
\item
Hull and White model - Table \ref{TabHW}.

As in \cite{ScarlattiAntonelli}, we have set: $t=0$, $T=0.5$, $S_0=100$, $v_0=0.2$, $r=0.0953$, $c=0.1$ and $\mu=0.2$.
Table \ref{TabHW} shows different percentage errors as $\rho$ and $K$ vary.
The percentage error is computed with respect to a benchmark value, obtained by evaluating the price of the European call option with a Monte Carlo method when the number of iteration is $10^6$ and the time interval is divided in $10^3$ parts.
\item
Stein and Stein model - Table \ref{TabSS}.

Again, the parameters are set as in \cite{ScarlattiAntonelli}, that is:
$t=0,\, T=0.5,\, S_0=100,\, v_0=0.2,\, r=0.0953,\, a=0.2,\, b=4, \,c=0.1$. The percentage errors are eveluated w.r.t. the exact price.

\item
Heston model - Table  \ref{HeparAlos} and \ref{HeparAS}.

The square-root model for the volatility process ensures that $v$ a.s. never reaches $0$ if $v_0\ge 0$ and the condition $2ab\ge c^2$ (often called Novikov condition) is fulfilled, and in such a case one has also nice Malliavin derivability properties for $v$ (see Al\acc os and Ewald \cite{Alos2}).  The results in \cite{ScarlattiAntonelli} are given when the Novikov condition does not hold. So, we present for this model two sets of results:
\begin{itemize}
\item
Table  \ref{HeparAlos} refers to parameters chosen in Al\`os \cite{Alos1}: $t=0$, $a=0.04$, $b=8$, $c=0.1$, $r=0.0953$, $v_0=0.0225$, $S_0=100$, so that condition $2ab\ge c^2$ is fulfilled;
\item
Table \ref{HeparAS} refers to parameters chosen in Antonelli and Scarlatti \cite{ScarlattiAntonelli}:  for short term options ($T=0.1$ or $T=0.2$) then $\rho=-0.76$, $a=0.025$, $b=1.62$ $c=0.44$; for long-term ones ($T=0.4$, $T=0.5$ or $T=0.8$) then $\rho=-0.64$, $a=0.035$, $b=1.15$, $c=0.39$.
Here, all cases give $2ab< c^2$.
\end{itemize}
The errors are computed w.r.t. the exact price, evaluated by using the pricer at \texttt{http:// kluge.in-chemnitz.de/tools/pricer/}.

\end{itemize}

It appears evident that Expansion A gives highly efficient results. In fact, the `ExpA-1' and `ExpA-2' percentage errors are always of order strictly smaller than $1\%$.
In addition, since `ExpA-1' coincides with the Antonelli and Scarlatti expansion up to the first order, it is also clear that often the approximations $\overline{g}_0$ and $\overline{g}_1$, giving a closed-form first order approximated formula, bring to loose accuracy. Moreover, the use of the second order coefficient generally improves the results (as shown in the `ExpA-2' errors), sometimes cutting the error of some order, and this is not really a problem from the numerical implementation point of view: the ExpA algorithm takes great advantage from the fact that the structures of the coefficients of order 1 and 2 are very similar each other, while the implementation of the second order coefficient from the Antonelli and Scarlatti expansion is actually heavy.

Concerning Expansion M, i.e. the one from the Malliavin approach,
%
in the Hull and White model and in the Stein and Stein one
the errors do not exceed $2\%$, so perfectly in the desired range. Such a performance also holds for the Heston model when the Novikov condition is fulfilled (Table \ref{HeparAlos}). But if such a condition does not hold (Table \ref{HeparAS}), the errors generally increase, even if they remain below the standard $5\%$ threshold but with two exceptions (ExpM-2: $\rho= - 0.64$, $K=110$ and ExpM-2: $\rho= - 0.76$, $K=110$), in which they are of order $6\%$. This could be explained by recalling that when the Novikov condition is not satisfied, the Malliavin differentiability of the volatility process is not guaranteed (see Al\`os and Ewald \cite{Alos2}).

Nevertheless, our opinion is that the largely quite accurate behavior of the Malliavin method is highly satisfying, mainly for two reasons. Firstly, as already observed, it is ready to be implemented for more general European options (even with a non regular payoff-function) and moreover, following Remark \ref{jumps}, it is worth to be used also if the log-returns are driven by the more general jump-diffusion model in (\ref{din-xi-jumps}).

\section{Tables}\label{sect-tables}

\begin{table}[!ht]
\centering
{\scriptsize{
\begin{tabular}{| l | l | l | r | r | r | r | }
\cline{3-7}
\multicolumn{2}{c|}{$\phantom{\displaystyle \sum}$}& $K=90$& $K=95$& $K=100$ & $K=105$ & $K=110$\\
\cline{1-7}
$\phantom{\displaystyle\sum}$& AS &  0.386235   &    0.440003  &  0.435233  &  0.339402  &  0.127030 \\
\cline{2-7}
$\phantom{\displaystyle \sum} $& ExpA-1 & 0.263909  & 0.309145	    & 0.353728	   & 0.395341	& 0.438145  \\
$\phantom{\displaystyle \sum}\rho=-0.25 $& ExpA-2 & 0.264210	& 0.309780	& 0.354730	& 0.396792	& 0.440186 \\
\cline{2-7}
$\phantom{\displaystyle \sum}$&ExpM-1   & 0.199430  & 0.425136	   & 0.469145	 & 0.725978  &0.910577 \\
$\phantom{\displaystyle \sum}$& ExpM-2  & 0.284079	& 0.496970	& 0.319980	& 0.267020	 & 0.514088  \\
\cline{1-7}
$\phantom{\displaystyle \sum}$& AS & 0.169833	& 0.141649	& 0.002633	& 0.335407	& 0.940766   \\
\cline{2-7}
$\phantom{\displaystyle \sum} $& ExpA-1  & 0.041942	& 0.080049	& 0.117349	& 0.152940	& 0.196648   \\
$\phantom{\displaystyle \sum}\rho=-0.5$
& ExpA-2  & 0.040742	& 0.077808	& 0.114099	& 0.148380	& 0.189953  \\
\cline{2-7}
$\phantom{\displaystyle \sum}$& ExpM-1   & 0.007453 & 0.124952	& 0.566645	& 0.239341 &0.315632 \\
$\phantom{\displaystyle \sum}$&ExpM-2  & 0.415920	& 0.529802	& 0.795568	& 0.518876	 & 1.636366    \\
\cline{1-7}
\end{tabular}
}}
\caption{{\small Hull and White model - percentage errors w.r.t. a benchmark value.
Parameters: $t=0$, $T=0.5$, $S_0=100$, $v_0=0.2$, $r=0.0953$, $c=0.1$, $\mu = 0.2$. \label{TabHW}}}
\end{table}

\begin{table}[!ht]
\centering
{\scriptsize
\begin{tabular}{|l| l | r | r | r | r | r | }
\cline{3-7}
\multicolumn{2}{c|}{$\phantom{\displaystyle \sum }$}& $K=90$& $K=95$& $ K=100$ & $K=105$ & $K=110$\\
\cline{1-7}
$\phantom{\displaystyle \sum }$& AS &   0.003941   &    0.123226  &  0.326447  &   0.583335 &    0.797076\\
\cline{2-7}
$\phantom{\displaystyle \sum }$& ExpA-1 & 0.007070	& 0.003160	& 0.011961	& 0.032119	& 0.039606 \\
$\phantom{\displaystyle \sum } \rho=-0.25$&
ExpA-2  & 0.005262	& 0.000415	& 0.014921	& 0.035616	& 0.046035 \\
\cline{2-7}
$\phantom{\displaystyle \sum }$&
ExpM-1   & 0.258058	& 0.240485	& 0.231560	 & 0.167886	& 0.149144\\
$\phantom{\displaystyle \sum }$&ExpM-2  & 0.163152	& 0.027884	& 0.069982	& 0.563581	 & 1.689356\\
\cline{1-7}
$\phantom{\displaystyle \sum }$& AS &  0.047087  &   0.160828 &  0.332411 &  0.571909  &  0.854295\\
\cline{2-7}
$\phantom{\displaystyle \sum }$& ExpA-1  & 0.051959	& 0.057347	& 0.024588	& 0.012997	& 0.035652 \\
$\phantom{\displaystyle \sum } \rho=-0.5$ &
ExpA-2 & 0.044712	& 0.045937	& 0.011695	& 0.028809	& 0.007137 \\
\cline{2-7}
$\phantom{\displaystyle \sum }$&
ExpM-1   & 0.336514	& 0.383521	& 0.449586	 & 0.293623 & 0.452388 \\
$\phantom{\displaystyle \sum }$&ExpM-2   & 0.295035	& 0.307504	& 0.506819	& 0.527940	 & 0.521331\\
\cline{1-7}
$\phantom{\displaystyle \sum }$& AS &   0.144578  & 0.239578   &   0.340785 &  0.522829 &  0.950355 \\
\cline{2-7}
$\phantom{\displaystyle \sum }$& ExpA-1 & 0.128902	& 0.157840	& 0.133673	& 0.106995	& 0.285850 \\
$\phantom{\displaystyle \sum }\rho=-0.75$&
ExpA-2 & 0.112629	& 0.130979	& 0.101239	& 0.065219	& 0.212845\\
\cline{2-7}
$\phantom{\displaystyle \sum }$&
ExpM-1   & 0.390693	& 0.164782	& 1.133769	 & 1.812543	& 1.634112\\
$\phantom{\displaystyle \sum }$&ExpM-2   & 1.650679	& 0.136951	& 1.524349	& 1.675562	 & 1.492334 \\
\cline{1-7}
\end{tabular}
\caption{{\small Stein and Stein model - percentage errors w.r.t. the exact price. Parameters: $t=0$, $T=0.5$, $S_0=100$, $v_0=0.2$, $r=0.0953$, $a=0.2$, $b=4$, $c=0.1$. \label{TabSS}}}
}
\end{table}

\begin{table}[!ht]
\centering
{\scriptsize
\begin{tabular}{|l |l| l | r | r | r | r | r | }
\cline{4-8}
\multicolumn{3}{c|}{$\phantom{\displaystyle \sum}$}& $K=90$& $K=95$& $K=100$ & $K=105$ & $K=110$\\
\cline{1-8}
$\phantom{\displaystyle \sum}$&$T=0.5$&ExpA-1  & 0.006135	& 0.007329	& 0.003350	& 0.002555	 & 0.002873\\
$\phantom{\displaystyle \sum}$&&
ExpA-2  & 0.005404	& 0.005785	& 0.001136	& 0.005671	& 0.007891\\
\cline{3-8}
$\phantom{\displaystyle \sum}$&&
ExpM-1   & 0.128272	& 0.122593	& 0.068726	& 0.135990	 & 0.060668	\\
$\phantom{\displaystyle \sum}$&&ExpM-2   & 0.663177 & 0.431397	& 1.190316	& 0.512402	 & 0.659557	\\
\cline{2-8}
$\phantom{\displaystyle \sum}$&$T=0.8$&ExpA-1  & 0.001454 & 0.001302 & 0.008102	& 0.018863	 & 0.031219 \\
$\phantom{\displaystyle \sum}\rho=-0.5$&&
ExpA-2  & 0.000532	& 0.002577	& 0.009491	& 0.020320	& 0.033123 \\
\cline{3-8}
$\phantom{\displaystyle \sum}$&&
ExpM-1  & 0.434218 & 0.500039 & 0.605644	& 0.373685	& 0.142499 \\
$\phantom{\displaystyle \sum}$&&ExpM-2  & 0.681888	& 0.145416	& 0.853767	& 0.912884	 & 0.473907\\
\cline{2-8}
$\phantom{\displaystyle \sum}$&$T=1$&ExpA-1 & 0.021982	& 0.031822	& 0.042954	& 0.055580	 & 0.071431	\\
$\phantom{\displaystyle \sum}$&&
ExpA-2  & 0.020960	& 0.029948	& 0.040171	& 0.051814	& 0.066535 \\
\cline{3-8}
$\phantom{\displaystyle \sum}$&&
ExpM-1   & 0.388991 & 0.553972	& 0.872142	& 0.734878	 & 0.642501\\
$\phantom{\displaystyle \sum}$&&ExpM-2   & 0.351164 & 1.212113	& 0.845210	& 0.171887	 & 1.006377\\
\cline{1-8}
\end{tabular}
\caption{{\small Heston model with parameters fulfilling the Novikov condition - percentage errors w.r.t. the exact price. Parameters: $t=0$, $S_0=100$, $v_0=0.02225$, $r=0.0953$, $a=0.04$, $b=8$, $c=0.1$. \label{HeparAlos}}}
}
\end{table}

\begin{table}[!ht]
\centering
{\scriptsize
\begin{tabular}{ |l |c| l | r | r | r | r | r | }
\cline{4-8}
\multicolumn{3}{c|}{$\phantom{\displaystyle \sum}$}& $K=90$& $K=95$& $K=100$ & $K=105$ & $K=110$\\
\cline{1-8}
$\phantom{\displaystyle \sum}$&&AS& 0.004745  & 0.052284 & 0.096813 & 0.149266  & 0.231650\\
\cline{3-8}
$\phantom{\displaystyle \sum}$&&ExpA-1 & 0.145056	& 0.084289	& 0.033238	& 0.052816	& 0.213478	\\
$\phantom{\displaystyle \sum}$&$T=0.4$&ExpA-2  & 0.076859	& 0.014733	& 0.041668	& 0.040969	& 0.078879\\
\cline{3-8}
$\phantom{\displaystyle \sum}$&&ExpM-1  & 0.049668	& 0.169833	& 0.040386	& 0.330853	 & 0.070622\\
$\phantom{\displaystyle \sum}$&&ExpM-2  & 1.682665	& 2.722891	& 1.967607	& 1.843184	 & 3.016751	
\\
\cline{2-8}
$\phantom{\displaystyle \sum}$&&AS&   0.004457 & 0.058372 &  0.109874 &  0.170933 & 0.264214\\
\cline{3-8}
$\phantom{\displaystyle \sum}$&&ExpA-1 & 0.217793	& 0.163739	& 0.124153	& 0.159125	& 0.339199\\
$\phantom{\displaystyle \sum}$&$T=0.5$&ExpA-2  & 0.121799	& 0.065143	& 0.018367	& 0.031568	& 0.166553\\
\cline{3-8}
$\phantom{\displaystyle \sum}\rho=-0.64$&&ExpM-1  & 0.295190 & 0.250897	& 0.348796	& 0.545558 & 0.301669 \\
$\phantom{\displaystyle \sum}$&&ExpM-2  & 0.956495	& 0.738435	& 3.272917	& 1.394025	 & 6.177427\\
\cline{2-8}
$\phantom{\displaystyle \sum}$&&AS&  0.023857  & 0.046118 &  0.115713 & 0.201686 & 0.326068\\
\cline{3-8}
$\phantom{\displaystyle \sum}$&&ExpA-1 & 0.040934	& 0.094668	& 0.230403	& 0.316294	& 0.292525	\\
$\phantom{\displaystyle \sum}$&$T=0.8$ &ExpA-2 & 0.134341	& 0.274656	& 0.420120	& 0.531201	& 0.557421\\
\cline{3-8}
$\phantom{\displaystyle \sum}$&&ExpM-1  & 0.269246	& 0.629545	& 1.258894	& 1.664970	 & 2.582807\\
$\phantom{\displaystyle \sum}$&&ExpM-2  & 2.030071	& 0.960147	& 2.772005	& 1.073997	 & 3.222600 \\
\cline{1-8}
$\phantom{\displaystyle \sum}$ &&AS& 0.044528  & 0.008180 & 0.051279 &  0.057639 & 0.051020\\
\cline{3-8}
$\phantom{\displaystyle \sum}$&&ExpA-1 & 0.075473	& 0.031628	& 0.034223	& 0.036352	& 0.479536\\
$\phantom{\displaystyle \sum}$&$T=0.1$&ExpA-2  & 0.061353	& 0.016046	& 0.051183	& 0.002745	& 0.399934	 \\
\cline{3-8}
$\phantom{\displaystyle \sum}$&& ExpM-1     & 0.715144	& 0.086117	& 0.484912	& 0.177757	 & 0.709826\\
$\phantom{\displaystyle \sum}\rho=-0.76$&&ExpM-2  & 1.690227  & 4.377866  & 0.592030	 & 4.588024	& 5.912330 \\
\cline{2-8}
$\phantom{\displaystyle \sum}$&&AS&  0.043600 & 0.029069 & 0.090594 &  0.121116 &0.082985\\
\cline{3-8}
$\phantom{\displaystyle \sum}$&&ExpA-1 & 0.138635	& 0.057408	& 0.030773	& 0.014847	& 0.395983\\
$\phantom{\displaystyle \sum}$&$T=0.2$&ExpA-2  & 0.094970	& 0.012563	& 0.079821	& 0.059268	& 0.259167	\\
\cline{3-8}
$\phantom{\displaystyle \sum}$&&ExpM-1  & 0.390806	& 0.121161	& 0.747368	& 1.999056	 & 1.871387\\
$\phantom{\displaystyle \sum}$&&ExpM-2  & 1.712282	& 0.612300	& 1.215959	& 1.571722	 & 2.452193\\
\cline{1-8}
\end{tabular}
\caption{{\small Heston model with parameters not fulfilling the Novikov condition - percentage errors w.r.t. the exact price. Parameters: $t=0$, $S_0=100$, $v_0=0.02225$, $r=0.0953$; for $\rho=-0.64$: $a=0.025$, $b=1.62$, $c=0.44$; for $\rho=-0.76$: $a=0.035$, $b=1.15$, $c=0.39$.
\label{HeparAS}}}
}
\end{table}

\clearpage

\textbf{Acknowledgments.} The authors would like to thank Elisa Al\`os and Antonino Zanette for their courtesy and for their pertinent and useful suggestions.




\end{document}